%% file: main.tex
\documentclass{acmart}

\AtBeginDocument{%
  \providecommand\BibTeX{{%
    \normalfont B\kern-0.5em{\scshape i\kern-0.25em b}\kern-0.8em\TeX}}}

\setcopyright{acmcopyright}
\copyrightyear{2018}
\acmYear{2018}
\acmDOI{XXXXXXX.XXXXXXX}

\acmJournal{TOCHI}
\acmVolume{37}
\acmNumber{4}
\acmMonth{8}

\input{utils}
\begin{document}

\title[Make It Make Sense!]{Make It Make Sense! Understanding and Facilitating Sensemaking in Computational Notebooks}


\author{Souti Chattopadhyay}
\email{schattop@usc.edu}
\affiliation{%
  \institution{University of Southern California}
  \state{California}
  \country{USA}
}

\author{Zixuan Feng}
\email{fengzi@oregonstate.edu}
\authornote{The second author contributed equally to this work as the first author.}

\affiliation{%
  \institution{Oregon State University}
  \state{Oregon}
  \country{USA}
}

\author{Emily Arteaga}
\email{arteagae@oregonstate.edu}
\affiliation{%
  \institution{Oregon State University}
  \state{Oregon}
  \country{USA}
}

\author{Audrey Au}
\email{auau@oregonstate.edu}
\affiliation{%
  \institution{Oregon State University}
  \state{Oregon}
  \country{USA}
}

\author{Gonzalo Ramos}
\email{goramos@microsoft.com}
\affiliation{%
  \institution{Microsoft Research}
  \state{Washington}
  \country{USA}
}

\author{Titus Barik}
\email{tbarik@microsoft.com}
\affiliation{%
  \institution{Microsoft}
  \state{California}
  \country{USA}
}

\author{Anita Sarma}
\email{anita.sarma@oregonstate.edu}
\affiliation{%
  \institution{Oregon State University}
  \state{Oregon}
  \country{USA}
}








\renewcommand{\shortauthors}{Chattopadhyay, et al.}

\begin{abstract}

Reusing and making sense of other scientists' computational notebooks. However, making sense of existing notebooks is a struggle, as these reference notebooks are often exploratory, have messy structures, include multiple alternatives, and have little explanation. 
To help mitigate these issues, we developed a catalog of cognitive tasks associated with the sensemaking process. Utilizing this catalog, we introduce \outliner: an interactive overlay on computational notebooks. \Outliner integrates computational notebook features with digital design, grouping cells into labeled sections that can be expanded, collapsed, or annotated for improved sensemaking.

We investigated data scientists’ needs with unfamiliar computational notebooks and investigated the impact of Porpoise adaptations on their comprehension process. Our counterbalanced study with 24 data scientists found Porpoise enhanced code comprehension, making the experience more akin to reading a book, with one participant describing it as \textsl{It's really like reading a book.}




\end{abstract}



\maketitle

\input{introduction.tex}

\label{sec_intro}

\input{sense_making}
\label{sec_research_setting}

\input{system_design.tex}

\label{system_design}

\input{design_probe}

\label{design_probe}

\input{user_evaluation}

\label{sec_user_evaluation}

\input{discussion} 
\label{sec_discussion}

\input{conclusion.tex}

\label{conclusion}

\section*{ACKNOWLEDGMENTS}
We thank all the interviewees for their contributions to this research. We would also like to express our gratitude to Marjan Adeli for her contribution to Appendix, which involved designing the classification of code functionality and functionality patterns. This work is supported by the National Science Foundation (NSF) under Grant Nos. CCF: 2008089 and IUSE: 2235601.

\bibliographystyle{IEEEtranN}
\bibliography{bib.bib}

\newpage

\section*{Appendix}

\subsection{Classification of Code Functionality}
\label{appen_classfication}

\begin{figure}[ht!]
\centering
\includegraphics[width=5 in]{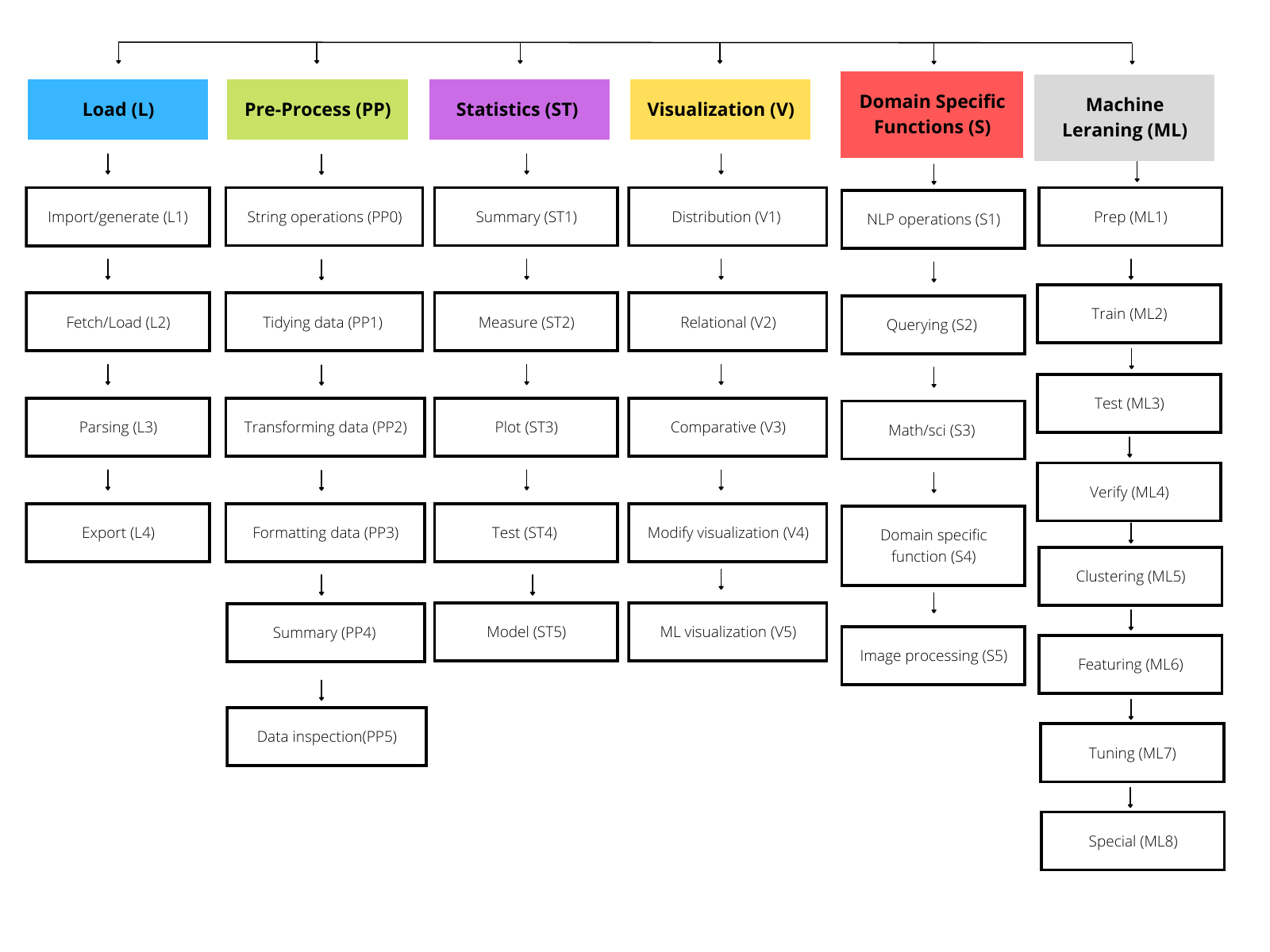}
\caption{Classification of Code Functionality (snippet): color cells indicate the category's name; grey cell shows examples of function calls under each category.}
\label{fig:ds_functions}
\Description{Figure 7 displays a classification of code functionality. Color-coded cells indicate the name of each category, while grey cells provide examples of function calls pertaining to each category. Starting from the top left corner, there's a rectangle labeled "Load (L)", symbolizing the data loading stage. Directly to its right, another rectangle is labeled "Pre-Process (PP)", which represents the data pre-processing stage. Adjacent to it is a rectangle marked "Statistics (ST)", denoting the stage of generating statistics from the data. To its right, a rectangle labeled "Visualization (V)" indicates the data visualization stage. Subsequently, there's a rectangle labeled "Domain Specific Functions (S)", embodying a set of functions specific to the project domain. Lastly, a rectangle labeled "Machine Learning Functions (ML)" represents a set of machine learning functions. For each category, it contains sub-codes that directionally connect with each other. For instance:
Under the "Load" category, there are four codes: Import/Generate (L1), Fetch/Load (L2), Parsing (L3), and Export (L4).
In the "Pre-Process" section, abbreviated as "PP", five sub-codes directionally connect: String Operations (PP0), Tidying Data (PP1), Transforming Data (PP2), Formatting Data (PP3), Summary (PP4), and Data Inspection (PP5).
For "Statistics", abbreviated as "ST", the sub-codes range from ST1 to ST5: Summary, Measure, Plot, Text, and Model.
In the "Visualization" category, shortened as "V", sub-codes range from V1 to V5, encompassing Distribution, Relational, Comparative, Modify Visualization, and ML Visualization.
Within "Domain Specific Function", sub-codes span from S1 to S5, covering NLP Operations, Querying, Math/Science, Domain Specific Functions, and Image Processing.
For the final category, "Machine Learning", abbreviated as "ML," sub-codes range from ML1 to ML8, which include Prep, Train, Test, Verify, Clustering, Featuring, Tuning, and Special.}
\end{figure}

\cref{fig:ds_functions} displays the functional categories of encoding transforms. Each line of the notebook is converted into a list ($H_1$ - $H_7$), as discussed in \cref{sec:3} of the paper. The 39 categories emerged from the first round of inductive coding between two researchers across all 35 notebooks and 605 function calls. For example, $pandas.read\_csv()$ is a function call that loads data and categorizes it as a `Load (L2)' functionality. Another example is $t.test()$, which we categorized as the `Statistical Test (ST4)' functionality. These categories naturally fell into seven broader categories: Load (L), Domain Specific Functions (S), Pre-Processing (PP), Visualization (V), Machine Learning (ML), Statistics (ST), and Others (O). 

The researchers met to negotiate their disagreements~\cite{forman2007qualitative} during inductive coding, which resulted in re-assigning categories to some functions and restructuring six categorizations (L3 $\longrightarrow$ PP5, S3 $\longrightarrow$ L3, S4 $\longrightarrow$ S3, S5 $\longrightarrow$ S4, S6 $\longrightarrow$ S5, S0 $\longrightarrow$ PP0) and removing the `Other' category as it wasn't indicating any meaningful grouping of functionality. The final categories of all observed functions are listed in \cref{fig:ds_functions}, and definitions of each category and functions belonging to them are provided in the supplementary materials~\cite{supp2023semsemake}.

\newpage
\subsection{Code Purpose from Functionality Patterns}
\label{appen_code_purpose}

\input{tables/Patterns_with_Icon}

To identify functionality patterns, for each line of code, we extracted the sequence of categories ($H_1$) based on their linear order.
Then, we looked at permutations and nested functions, for instance, $np.mean (accuracy (y_{test}, pred)$ performs a summary (ST1) on the output of a machine learning model (ML4). We defined the rule to infer processes from nested calls in execution order($H_2$, $H_3$). We can transform notebook code into a list of categories using these three heuristics ($H_1$ - $H_3$).
Loading data (category L) in any form indicates the start of a part of code written for a specific purpose ($H_4$), and getting any form of output like visualizations (V) or of a model (ML4) indicates the end of a purpose ($H_5$). Repeated subsequences were investigated to identify meaning ($H_6$). Finally, if adjacent lines of code performed the same function, they were all grouped into one process, and the sequence was instead counted as 1:PP3 ($H_7$). Using these heuristics ($H_4$ - $H_7$), we process the notebooks into an encoded list of categories that reflect the `purpose' or functionality of the code. \cref{fig:ds_functions} lists all the patterns in functionally we identified across all the notebooks. 
For a detailed description of the process of the code purpose, please refer to the supplementary material~\cite{supp2023semsemake}. 

\end{document}

%% file: utils.tex
\usepackage{tabularx}
\usepackage{graphicx}
\usepackage{textcomp}
\usepackage[ampersand]{easylist}
\usepackage{multicol}
\usepackage{wrapfig}

\usepackage{multirow}
\usepackage{fontawesome}

\usepackage{threeparttable}
\usepackage{tabularx}

\usepackage{rotating}
\usepackage[roman]{parnotes}
\usepackage[many]{tcolorbox}
\usepackage{soul}
\usepackage{xcolor}
\usepackage{comment}
\usepackage{booktabs}
\usepackage{colortbl}
\usepackage{verbatim}
\usepackage{xspace}
\usepackage{xcolor}
\xspaceaddexceptions{]\}}

\usepackage{hyperref}
\usepackage[capitalize,noabbrev]{cleveref}
\usepackage{footmisc}
\usepackage{array}
\usepackage[inline]{enumitem}
\usepackage[export]{adjustbox}
\usepackage{longfbox}
\usepackage{commath}
\usepackage[normalem]{ulem}
\usepackage{mdframed,lipsum}
\usepackage{ifdraft}
\usepackage{wrapfig}
\usepackage{graphicx}
\usepackage{makecell}
\usepackage{quoting}
\usepackage{fontawesome}

\newif\ifdraft
\drafttrue 

\usetikzlibrary{calc}

\definecolor{labelColor}{RGB}{56,152,236}
\newcommand{\overviewLabel}[1]{\tikz [font=\sffamily, baseline={($ (current bounding box.center) - (0,.3em) $)}] 
\fill[fill=labelColor] (0,0) circle (0.5em) node[text=white] {#1};}

\definecolor{labelColorBig}{RGB}{90, 79, 207}
\newcommand{\overviewLabelBig}[1]{\tikz [font=\sffamily, baseline={($ (current bounding box.center) - (0,.3em) $)}] \fill[fill=labelColorBig] (0,0) circle (0.5em) node[text=white] {#1};}

\newcommand{\outliner}{\textsc{Porpoise}\xspace}
\newcommand{\Outliner}{\textsc{Porpoise}\xspace}

\newcommand{\changed}[1]{\textcolor{black}{#1}}
\newcommand{\reorg}[1]{\textcolor{black}{#1}}

%% file: introduction.tex
\section{Introduction}

Sharing knowledge is an inevitable part of day-to-day organizational life; building software and information artifacts of substantial size requires the expertise and effort of multiple people~\cite{whitehead2010collaborative, Whitehead2010}.  In addition to creating computational notebooks, data scientists also write and disseminate their analytical artifacts among a wide range of stakeholders such as developers, program managers, and marketing specialists~\cite{wang2019data, kim2017data, subramanian2020casual, zhang2020data}. However, making sense of these artifacts, like computational notebooks, requires significant effort to comprehend the code and intentions behind those lines of code.


Data scientists frequently engage in exploratory programming, mixing and matching different solutions from their experiments or adopting approaches from other scientists' work, predominantly utilizing computational notebooks as their tool of choice~\cite{kery2018story, kery2017exploring, kery2017variolite, Rule2018Explore, painpoints2020chattopadhyay, Perkel2018}. These notebooks are particularly well-suited for exploratory processes, as they adhere to the principles of literate programming, aiming to enhance code comprehension by effectively incorporating the programmer's commentary within the code itself~\cite{knuth1984literate}. Notebooks allow data scientists to compile code cells, their commentary (as markdown text), and the output of these cells into a single, cohesive document~\cite{Rule2018folding,messes2019head}.

In an ideal situation where notebooks are simple, well-structured, and documented, data scientists can quickly understand the code in unfamiliar notebooks. But the reality is far different. Notebooks are complex (lengthy with complicated, custom-made functions)~\cite{kandel2012enterprise}, messy (multiple additional analyses co-exist, some obsolete, some erroneous)~\cite{Rule2018Explore, kery2018story}, and confusing (code cells don't have documentation or obvious cues about the purpose of the code)~\cite{pimentel2019large, head2019managing}. This makes it difficult for data scientists reviewing unfamiliar notebooks to understand the purpose.


To bridge this gap, we investigate the sensemaking process and identify how to support sensemaking steps in computational notebooks. Leveraging the affordances that assist sensemaking in digital media, such as those found in e-books, we developed a design probe, \Outliner. This tool forms an interactive overlay on computational notebooks (for example, Jupyter Notebook), enhancing the user's ability to navigate and comprehend the notebook by providing a structured and easy-to-follow layout, much like the chapter and section divisions in e-books. It achieves this by drawing out and spotlighting comments on the underlying rationale of the code, thus offering users a more straightforward, more coherent narrative of the computational processes involved. We structured our research around these two questions:

\begin{itemize}
    \item RQ1: How can we design computational notebooks to help sensemaking?
    \item RQ2: How helpful are computational notebooks that support sensemaking?
\end{itemize}

\begin{figure}
\centering
\includegraphics[width=4.5 in]{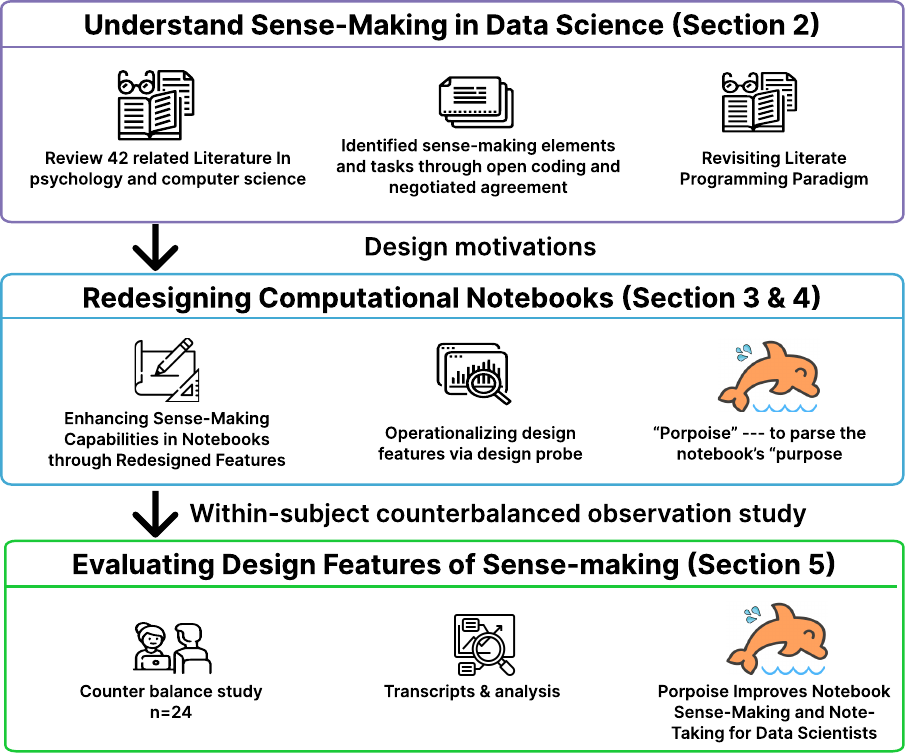}
\caption{Study Overview. We study sensemaking literature across domains in \cref{sec:background} to form design motivations for our features in \cref{sec:3} and build \outliner in \cref{design_probe}. Finally, we evaluate \Outliner in \cref{sec:sec5} using a within-subject study with 24 particpipants}
\Description{Figure 1 presents a methodological diagram illustrating the three phases of this study. The diagram is segmented into three main sections, each denoting a distinct research phase.
The topmost section, shaded in purple, is labeled "Understand Sense-Making in Data Science (Section 2 of the Paper)". Within this section are three sub-sections: "Review of 42 Related Literature in Psychology and Computer Science", "Identification of Sense-Making Elements and Tasks via Open Coding and Negotiated Agreement", and "Revisiting the Literate Paradigm". These sub-sections encapsulate the preliminary stages of the research and lay the groundwork for the subsequent study.
The middle section, tinted in blue, bears the label "Redesigning Computational Notebooks (Sections 3 & 4 of the Paper)". This section focuses on enhancing sense-making capabilities in notebooks. It consists of sub-sections such as "Enhancing Sense-Making Capabilities in Notebooks through redesigned features," "Operationalizing design features via design probe," and "Porpoise to parse the notebook’s purpose." These sub-sections highlight the efforts to redesign computational notebooks to improve sense-making in data science.
The lower section is labeled "Evaluating Design Features of Sense-Making (Section 5 of the Paper), highlighted in green. This section focuses on evaluating the design features related to sense-making. It includes sub-sections like "Counterbalance study", "Transcripts & analysis",  and "results shows that Porpoise Improves Notebook Sense-Making and Note-Taking for Data Scientists." These sub-sections represent the evaluation and analysis phase of the research project.
In summary, the diagram offers a comprehensive snapshot of the various phases and elements intrinsic to our research on sense-making within data science.
}
\label{fig:overview}
\vspace{-5mm}
\end{figure}

Answering research questions, we followed a systematic approach as shown in \cref{fig:overview}. 
We began by reviewing the literature on the elements of sensemaking across psychology, computer science, and data science. We identified three common elements: comprehension~\cite{woods1997automation, von1993code, bryant2011software, wang2003cognitive,ntuen2006cognitive, klein2006making, snowden2011naturalizing, snowden1999story, borges2010learning, fiol1985organizational, wang2006layered}, mental modelling~\cite{johnson1989mental, rutherford1991models, rouse1986looking, moray1998identifying,abel1998mental, johnson1989mental, langan2004mental, vosniadou199416, nersessian2002cognitive}, and situational awareness~\cite{tiemens1989cognitive, klein2007data, endsley2017toward,bryant2011software, klein2006making, endsley2017toward}.

We then developed design features to enhance sensemaking in data science notebooks using \Outliner in \cref{sec:3}. We then operationalized our design features for \Outliner as an interactive overlay for Jupyter notebooks in \cref{design_probe}. To evaluate the effectiveness of \Outliner and how it could address data scientists' sensemaking needs, we conducted a within-subject counterbalanced observation study with 24 data science practitioners.

Our results showed that participants using \Outliner were enthusiastic about the affordances \Outliner provides to the navigation and sensemaking of notebooks. \Outliner provided a gradual information overlay that allowed them to progressively build insights and inferences as they made sense of the notebook. Moreover, participants found that \outliner helped them find interesting and important areas in the notebook, share insights, and receive input from others. Our findings could inform future researchers and tool builders about the underlying cognitive tasks of sensemaking to inspire their design choices for future implementations of notebooks.

%% file: sense_making.tex
\section{Sensemaking in Data Science}
\label{sec:background}

\subsection{Literature Review about Sensemaking}

To support data scientists in their cognitive tasks within computational notebooks, it is essential to understand their needs. Our approach began with a review of related literature on sensemaking in psychology. Then, we expanded our scope to include the terms:``Computer Science'' and ``Data Science'', to provide a more comprehensive understanding of sensemaking in these scopes. Additionally, we reviewed digital books and media content to investigate how their features enhance the comprehensibility of books and media content.

We conducted a pilot search on Google Scholar by which we found relevant terms used in sensemaking, particularly in psychology. The final list of search keywords included ``sensemaking,'' ``sense making'', ``sensemaking in data science'', ``sensemaking in computer science'', ``comprehending activities'', and ``sensemaking psychology'', yielded 42 articles.

After identifying relevant literature by the initial screening process, we selected 15 papers that discussed sensemaking in the abstract. Then, the first and second authors followed the guidelines recommended by \citet{keele2007guidelines} and performed a single iteration backward snowballing~\cite{wohlin2014guidelines} to identify additional research on the topic. This led us to six additional studies related to education and computer science literature that provided detailed insights into the elements of sensemaking for computer science and data science, bringing the total number of papers in our final list to 21.

The first two authors independently read and analyzed the 21 papers following the open coding protocol~\cite{glaser2016open}. We held weekly meetings to present our findings and discuss them until we reached an agreement. After the meetings, we agreed on classifying sensemaking into three main categories, each encompassing the elements of sensemaking, their definitions, and the cognitive tasks associated with each element. Through this analysis, we comprehensively understood the elements of sensemaking, their definitions, and the associated cognitive tasks.

\begin{figure}
\centering
\includegraphics[width=5 in]{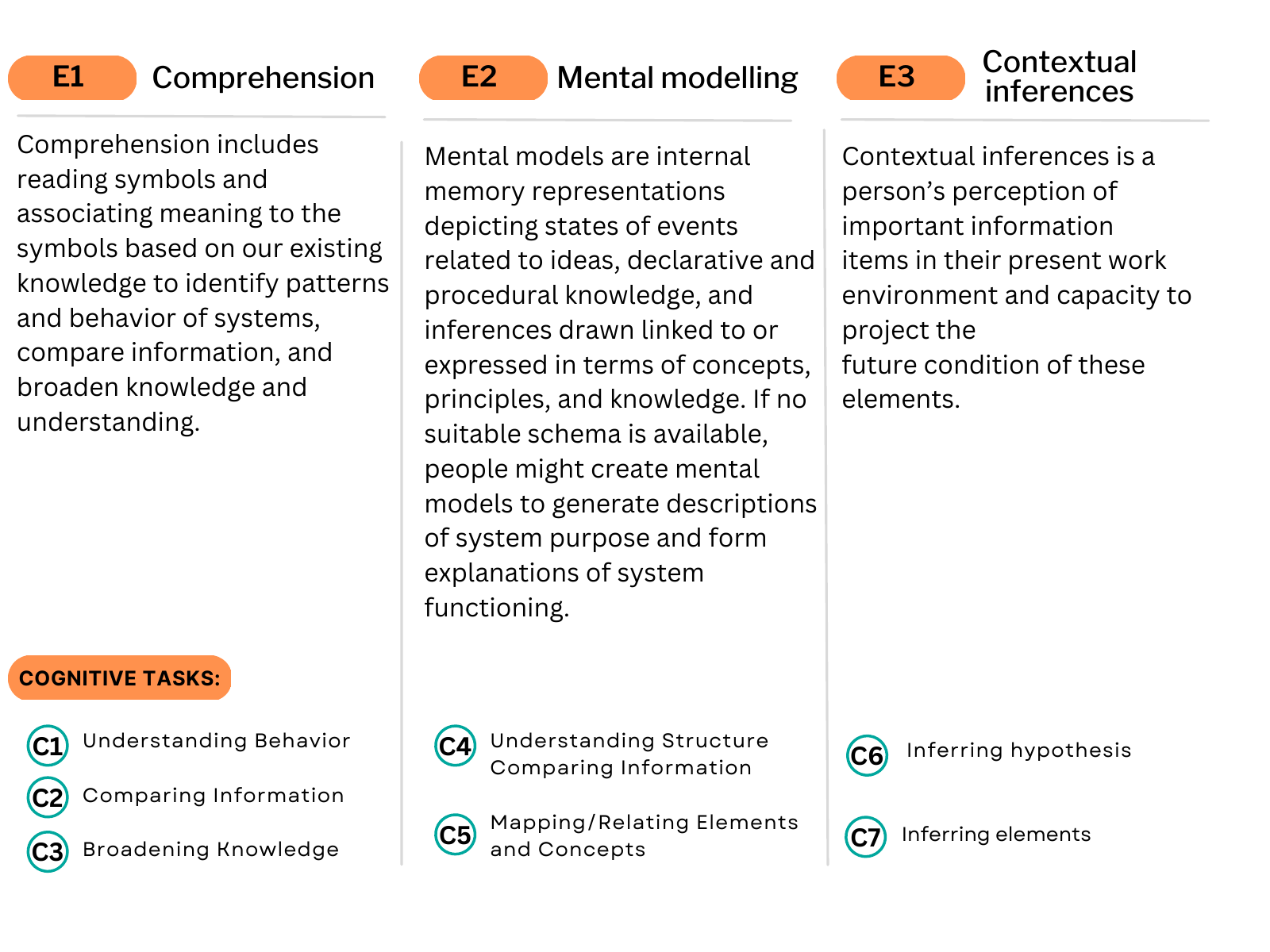}
\caption{Defining Sensemaking and Its Related Cognitive Tasks. The sensemaking process has three elements: E1. Comprehension, E2. Mental Modelling, and E3. contextualizing inferences. Each element is associated with a cognitive task listed below.}
\label{fig:sense_making}
\vspace{-5mm}
\Description{The figure 2 illustrates our definition of sensemaking and its associated cognitive tasks. The sensemaking process comprises three main components: E1 (Comprehension), E2 (Mental Modelling), and E3 (Contextualizing Inferences). Each of these components corresponds to different cognitive tasks.

On the far left, the E1 component is represented by the text "Comprehension." The definition beneath this title reads, “Comprehension involves interpreting symbols and attributing meaning to them based on our existing knowledge. This allows us to identify patterns and behaviors of systems, compare information, and expand our knowledge and understanding.” The cognitive tasks associated with E1 include "Understanding Behavior" (C1), "Comparing Information" (C2), and "Broadening Knowledge" (C3).

In the center, the E2 component is denoted by the text "Mental Modelling." It's explained that mental models are internal memory representations that depict event states related to ideas, both declarative and procedural knowledge, and inferences linked or expressed through concepts and principles. If there isn't a suitable schema available, individuals may create mental models to describe a system's intent and provide explanations for its operations. This component encompasses two cognitive tasks: "Understanding Structure" (C4) and "Elements and Concepts" (C5).

On the far right, the E3 component is represented by "Contextual Inferences." The definition states that contextual inference refers to an individual's recognition of key information items in their current work environment and their ability to anticipate the future states of these elements. The cognitive tasks related to this component are "Inferring Hypotheses" (C6) and "Inferring Elements" (C7).}
\end{figure}

\textbf{Sensemaking} is an integral part of any creative process \cite{drazin1999multilevel}. It is a complex activity that requires finding information, mapping the meaning to the larger context, and creating interpretations \cite{marchionini2019search}. When complexity in collaborative environments increases, sensemaking develops an artifact or solution \cite{wang2014framework}. Collaboration then involves iterative sensemaking of another person's program and adding knowledge this \cite{zhang2014towards}.

\textbf{The process of sensemaking} is a widely-discussed psychological perspective encompassing various elements. While some researchers associate sensemaking with comprehension alone~\cite{klein2006making, fitzgerald2019teaching}, others argue that it also entails mapping comprehended information to existing knowledge~\cite{klein2007data}. The sensemaking process involves integrating conjectured information with known data, linking an individual's inferences to their observations, elucidating ambiguous data, diagnosing uncertain symptoms, and pinpointing problems \cite{klein2007data}. In line with this, \citet{pirolli2005sensemaking} emphasizes that the general sensemaking process includes gathering information, building a schema, and creating insight. These aspects align with the three elements discussed in \cref{fig:sense_making}.

\textbf{Sensemaking in the context of programming} has been studied, but its application and implications for data scientists are still poorly understood. In the early 2000s, \citet{pirolli2005sensemaking} investigated the meaning of sensemaking in programming and found that programmers followed a sensemaking loop when building programs. Researchers have observed that programmers constructed narratives when attempting to comprehend someone else's code \cite{srinivasa2016foraging}, while frequently writing disorganized and ad-hoc code during exploration~\cite{kery2017variolite}. The computational notebook environment doesn't mandate a specific structure, and making code comments or Markdown explanations was optional and uncommon among data scientists~\cite{kery2018story}. A typical data science workflow, as described by \citet{kandel2011wrangler}, involves context-switching between raw data, wrangling tools, and visualization tools, suggesting an ``ideal'' solution would integrate these workflows into a single tool.

\textbf{E1. Comprehension} Different definitions of the three sensemaking elements include different cognitive tasks. For instance, understanding behavior~\cite{woods1997automation, von1993code, bryant2011software, wang2003cognitive}, comparing information~\cite{ntuen2006cognitive, klein2006making}, and expanding knowledge~\cite{snowden2011naturalizing, snowden1999story, borges2010learning, fiol1985organizational, wang2006layered} are low-level cognition-related tasks associated with the process of comprehension. Comprehension refers to an individual's awareness of facts important to the task and the broader purpose \cite{ntuen2006cognitive}, which includes reasoning, pattern recognition, and information comparison~\cite{klein2006making}. Combining these includes reading symbols and associating meaning to the symbols based on our existing knowledge to identify patterns and behavior of systems, compare information, and broaden knowledge and understanding.

\textbf{E2. Mental models} are internal memory representations that depict states of events related to ideas, declarative and procedural knowledge~\cite{johnson1989mental, rutherford1991models, rouse1986looking, moray1998identifying}, and inferences drawn connected to or expressed in terms of concepts, principles, and knowledge~\cite{abel1998mental, johnson1989mental, langan2004mental, vosniadou199416, nersessian2002cognitive}. When no suitable schema is available, individuals might create mental models to generate descriptions of system purposes and formulate explanations of system functioning.

\textbf{E3. Situational awareness} is defined as a person's perception of essential information items in their current work environment and their ability to project the future condition of these elements. The sensemaking process for information and structure involves generating hypotheses to explain discrepancies, searching for evidence to establish, support, or refute the hypotheses~\cite{tiemens1989cognitive, klein2007data, endsley2017toward}, and seeking a structure to integrate all discovered information~\cite{bryant2011software, klein2006making, endsley2017toward}.

\subsection{Revisiting Literate Programming Paradigm}

To design a computational notebook that supports this process and enhances data scientists' ability to engage in sensemaking, we must incorporate additional features that specifically support the cognitive tasks underlying sensemaking.

We then drew inspiration from the original literate programming paradigm behind computational notebooks~\cite{knuth1984literate}, which suggested that programs written as literary works are more understandable by humans. We also investigated features used in digital literature to make books and media more comprehensible. By combining insights from psychology and computer science, we developed a scaffolded notebook that provides users with a more structured and intuitive interface to empower them to engage in more effective sensemaking.

\textbf{Literate programming:} 
In the early 1980s, \citet{knuth1984literate} proposed a literate programming paradigm to make programming more explainable to human beings by writing it as a flow of thought rather than instructions to a computer. The earlier versions of literate programming interweaved code with programmers' explanations to create printable documents~\cite{Knuth2009Literate}. Recently,  computational notebook environments have adapted literate programming, e.g., Sage Notebooks~\cite {2005Sign}, \citet{Walter2019JupyterLab}, \citet{jupyter}. A computational notebook can have "cells" that include code, output, table, visualizations, etc. Data scientists can manually interweave their explanations in markdown cells~\cite{kery2018story}. Such computational notebooks were intended to help create and share computational narratives.

\textbf{Applying literate programming:} However, data analysis is iterative and exploratory~\cite{Rule2018Explore}, which creates messy computational notebooks with many erroneous code cells and throwaways~\cite{kandel2012enterprise, guo2012burrito, messes2019head}. Data scientists don't want to manually add explanations to these exploratory notebooks as there is no assurance the analysis will be used~\cite{Rule2018Explore}, which causes notebooks to stray away from the literate programming paradigm. \Outliner bridges this gap by grouping code cells logically and automatically annotating the groups with a description of the code's purpose.

%% file: system_design.tex
\section{DESIGNING COMPUTATIONAL NOTEBOOKS TO SUPPORT SENSEMAKING}
\label{sec:3}
 
To enhance data scientists' ability to engage in sensemaking within computational notebooks, we aim to re-design the features that support these cognitive tasks for data scientists. In this section, we seek to the
research question:

\textbf{RQ1: How can we design computational notebooks to help sensemaking?}

\subsection{Mapping features from Digital literature into Computational Notebooks}
\label{sec:feature-map}

We aim to build a computational notebook design with simple affordances that blend how people interact with digital and computational books. These features were proposed by the author's experiences with computational notebooks and HCI design, insights from prior research reporting data scientists' challenges with notebooks~\cite{painpoints2020chattopadhyay}, and affordances found in related systems or books. These adaptations of a set of simple affordances:

\changed{\begin{itemize}
    \item \overviewLabelBig{A} Navigation panel:  Provides an overview of the notebook's contents and structure through a side panel.  
    \item \overviewLabelBig{B} Annotation: Enables data scientists to create annotations by highlighting text.  
    \item \overviewLabelBig{C} Export: Allows users to create a snapshot of expanded chapters/sections by exporting a document.  
    \item \overviewLabelBig{D} Chapter title and icon: provide an introductory overview of the chapter and its sections. 
    \item \overviewLabelBig{E} Section title: As section headers that can be expanded to reveal their contents.
\end{itemize}}

\input{tables/design_mapping}

Table \ref{table:mapping} shows the relationship between cognitive tasks and design features. The table highlights seven cognitive tasks (C1 to C7) and their corresponding design features, marked as \overviewLabelBig{A} through \overviewLabelBig{E}. 



\begin{figure}
\centering
\includegraphics[width=5.5in]{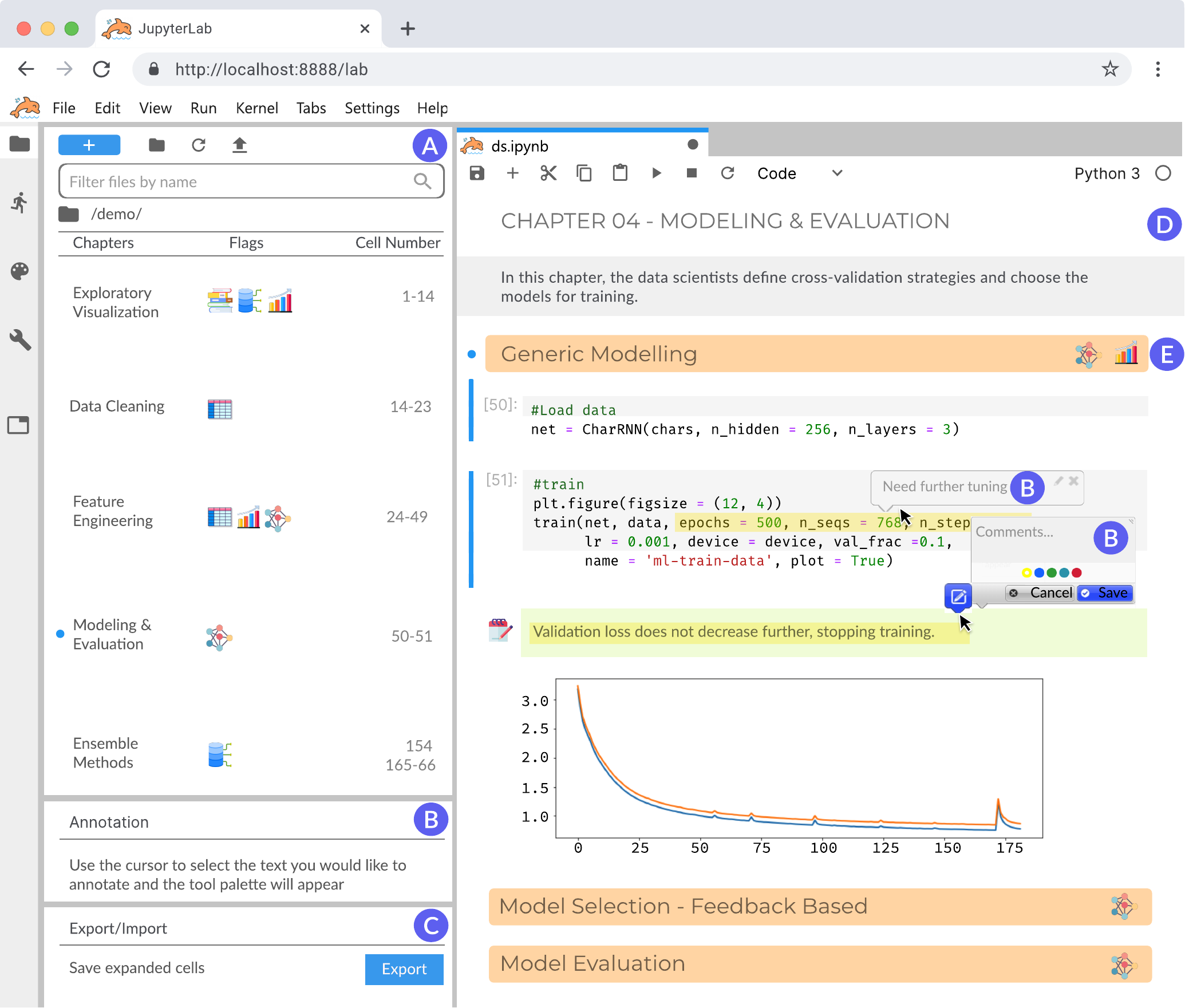}
\caption{\Outliner is an overlay interaction for computational notebooks that groups adjacent code cells automatically and conveys their purpose through five main interactive affordances.}
\label{fig:birdview}
\Description{The Figure 4 is a screenshot of the porpoise displaying in detail the feature description. In the top left of the image, there is a graph composed of smaller graphs, representing a small-scale version of the entire PORPOISE interface. A magnifier then zooms in to focus on both the left side panel and the main code sections on the right. Below the small-scale plot, there is a zoomed-in view of the left side panel, which displays the panel in detail. This panel functions similarly to a table of contents. It provides a streamlined overview by showcasing chapter titles, allowing users to quickly identify major chapters such as "exploratory visualization," labeled as number 1. Enhancing its utility, the side panel indicates the number of cells associated with each chapter marked as number 3. As shown in the figure, the Ensemble method's code spans 1-14 cells. The side panel also features "flags" labeled as number 2 that highlight key operations within the notebook, such as the importing of data and libraries, the display of graphs and tables, and the building of models. Additionally, another feature displays annotations labeled as number 4. Below the annotation, a description acts as a description, stating, "Use the cursor to select the text you wish to annotate, and the tool palette will appear." Further below, there's a feature labeled as number 5, titled "export/import." The description under this title indicates "Save expanded cells," along with a blue export button. On the right side of the small-scale plot, there is a zoomed-in view of the right expanding area. In the expanded Chapter 4 (Modeling & Evaluation), the content unfolds on the right-side canvas. This view displays the chapter title, labeled as number 1, followed by a brief introductory description of the chapter, labeled as number 6. The description reads, “In this chapter, data scientists define cross-validation strategies and select models for training.” Below the chapter description, there's a section title labeled as number 7. Within each section headers area, "flags" are presented in the section headers, labeled as number 2. Below the section header, there are two code boxes. The first box is commented as 'load data', and the second box is labeled 'train'. Under these code boxes, there is a markdown comment. At the very bottom, there is a plot generated by the 'train' code. Within the 'train' code, there's a dialogue bubble labeled as number 8, indicating existing comments, which read: “needs further tuning.” Additionally, a cursor signifies that a user is currently commenting on the markdown sentence labeled as number 9, allowing the user to leave comments. This feature provides different color options for highlighting comments.}
\end{figure}

Here, we present how we designed each feature based on existing literature and our experiences:
\textbf{\overviewLabelBig{A} Navigation panel.}  A common feature of digitized books is the table of contents, typically on the left side of the main text area \cite{PDF}. Visual representations of program segments rather than textual ones resulted in more quick and accurate program understanding \cite{cunniff1987graphics}. 
A data science project's lifetime may be divided into many stages, such as data acquisition, data readiness, etc \cite{wang2021autods}. 
Stakeholders and data scientists prefer presentation decks that are easily understood and visually represent high-level insights. \cite{kross2021orienting}. 
Because of these findings, we chose to concentrate on the navigational structure as one of the primary design affordances of \Outliner.

The content table provides a book's entire structure and layout at a glance, making it easy to get an overview that helps place the contents in context~\cite{wu2013table}. A common configuration of this table is showing the chapters' (or sections) titles and the associated page numbers~\cite{lin1997logical}. Indexes, or page numbers, indicate the amount of information contained in each chapter~\cite{o1997comparison}.
To provide an overview of a notebook's contents, we added a side panel displaying its major sections. The left panel of the notebook shows the chapter title and code cells included in the chapter. For example, in \cref{fig:birdview}, the chapter navigation panel \overviewLabelBig{A} shows the chapter ``Exploratory Visualization'' runs from cell 1 through 14, while the last chapter, Ensemble Methods include cell 154, and cells 165 and 166.

\changed{\textbf{\overviewLabelBig{B} Ability to take notes in notebook.} Taking notes while reading a physical book, online, or PDF articles is a very common activity supported by most digital media forms. It allows readers to go through the documents to commit information to memory and remind themselves of past thoughts without editing the information~\cite{kalyanpurthreaded}. While the most common way of allowing annotations is through comments and memos (also known as sticky) notes~\cite{ball2020sticky}, different characteristics of these annotation functions offer specific support. For instance, using color and highlighting in the text to improve reader comprehension by increasing reader engagement \cite{guernsey2015pioneering}. Annotations should also prevent covering over the raw data or parts of the text~\cite{o1997comparison} and should be able to contain information in a compact and flexible format~\cite{ball2020sticky}.}

We adapted these findings to design the annotation feature as a compact, inexpensive, rectangular, lightweight, flexible, and intuitive media for recording words, thoughts, drawings, and models \cite{ball2020sticky,garciadiseno} data scientists wish to use. At the same time, they go through unfamiliar notebooks to remind themselves of future references without having to edit the information \cite{kalyanpurthreaded}. The annotations feature, marked as \overviewLabelBig{B} in \cref{fig:birdview}, shows the characteristics of the annotation features. Data scientists can highlight any part of the notebook to annotate, which pops up a widget to write their thoughts. To enable data scientists to post comments and highlight text with different colors, we included various color options for the highlight so that users may use multiple colors to emphasize different contexts for different purposes. To prevent covering over the raw data by annotations such as sticky notes, we introduced the functionality that the annotated comments will only appear if the user hovers their cursor over the highlighted portion of the data~\cite{o1997comparison}.

\reorg{\textbf{\overviewLabelBig{C} Improving shareability of computational narratives of choice.} At the end of reading and making sense of a book and article, readers share or present their knowledge~\cite{pirolli2005sensemaking}. Sharing relevant information and the explanation of the information after sensemaking with coworkers improves the productivity of the team. Readers share digital media but email or host articles through the web. However, data scientists struggle with sharing notebooks, or parts of the notebook, easily with co-workers~\cite{painpoints2020chattopadhyay}, as it involves replicating packages, importing data and libraries, managing dependencies, reading and understanding someone's explanations of the code or responding to them. Additionally, computational notebooks have multiple narratives depending on the ordering and inclusion of cells.
To address these issues, we adapted the export feature marked as \overviewLabelBig{C} in \cref{fig:birdview}. The export feature exports only the parts of the notebook that are expanded and the contents within the expanded parts (including any annotations).}

\textbf{\overviewLabelBig{D}  \overviewLabelBig{E} Chapters, sections, and icons to provide structural and logical understanding.} To implement this navigation panel, we also need to identify within notebooks the set of structural features found in books and articles: chapters, sections and subsections. The information about the hierarchy of logical components, such as titles, abstracts, and sections, along with the physical components like pages, paragraphs, tables, figures, is useful in finding relevant information within textbook~\cite{mao2003document, gao2011structure} and conveys the underlying logical structure of the data in the article or book~\cite{bart2010information}. We adapted the notebook interface to elucidate similar logical and physical components.

Figure~\ref{fig:birdview}, chapter \overviewLabelBig{D} consists of the number, title, and a short description of the contents of the chapter akin to the introductory paragraph of a book or abstract of an article~\cite{sharp1995dynamic}. The sections within each chapter, marked as Figure~\ref{fig:birdview}\overviewLabelBig{E} , appear as highlighted bars containing explanations of the behavior of a group of code cells.
These bars are interactive, expanding to show code, output, and comments when clicked once and collapsing when clicked twice.

Similar to how physical components of books are defined as words, tables, figures~\cite{gao2011structure}, we identified the components of computational notebooks are data, libraries, visualizations, tables, models, and author notes. We displayed this information through a set of six icons that explain what the icon stands for when hovered over. This creates an affordance to forage for the relevant information, for instance, when a data scientist is looking for how specific models are defined by identifying which parts of the notebook have those components from the navigation panel and section headers. 

The combined ability to navigate around sections and chapters along with getting high-level structural information is shown to aid comprehension in digital media~\cite{chong2009design}. This structured overview can help data scientists create a mental map of the knowledge contained in the notebook~\cite{ran2020design}. It also eliminates the need to scroll for a long time, which is associated with a sharp decrease in attention~\cite{nnscrolling} and causes readers to lose their place in the document.

\subsection{Operationalizing Design Features}

To implement the design features introduced above from \overviewLabelBig{A} to \overviewLabelBig{E}, we need to identify chapters and sections in notebooks for the navigation panel \overviewLabelBig{A}, the chapters \overviewLabelBig{D}, and the sections \overviewLabelBig{E}. Therefore, before implementing these features, we must conduct additional analysis to define and automatically identify the chapters and sections in notebooks. However, it should be noted that implementing \overviewLabelBig{B} and \overviewLabelBig{E} is straightforward regarding user interaction design and thus doesn't require this additional analysis.

We defined the chapters to indicate broader sections of the analysis similar to steps in machine learning workflow like: ``Data Processing'' and ``Modelling''. The sections indicated the purpose of a group of consecutive (or one) code cells. However, how can we define and extract chapters and sections? How do data scientists structure their analyses into sections similar to chapters and headers? To answer these questions, we took a bottom-up approach by qualitatively analyzing existing notebooks from Open Source such as Jupyter Notebooks and Google Colaboratory.

\textbf{Qualitative analysis to identify notebook functionality.} We selected 100 notebooks hosted on GitHub for competitions in the Kaggle platform, the world's largest data science community \cite{Kaggle}. These notebooks were related to domains like climate, sports, finance, traffic, and viruses with executed output. After removing incomplete notebooks or notebooks that are drastically different from previously studied practitioner's notebooks \cite{messes2019head}, we arrived at a set of 46 notebooks, of which two were not executable. Finally, we analyzed 44 notebooks from our set to understand how to define and extract chapters and sections. We will describe this analysis in the paragraphs below.

Since chapters and sections are designed to capture the purpose of the code, we looked at the difference in functionality across the 44 notebooks. Two researchers classified every line of code (LOC) or function call in nine random notebooks (20\% of 44 rounded up) based on their functionality following an open inductive coding approach. For example, $pandas.read\_csv()$ is a function call that loads data and categorizes it as a `Load (L2)' functionality. Another example is $t.test()$, which we categorized as the `Statistical Test (ST4)' functionality. After the first round of inductive coding~\cite{Saldana2009}, 39 such categories emerged. These categories naturally fell into seven broader categories: Load (L), Domain Specific Functions (S), Pre-Processing (PP), Visualization (V), Machine Learning (ML), Statistics (ST), and Others (O). The two researchers independently coded the LOCs/function calls across 35 notebooks. Across all notebooks, we categorized 605 different function calls. The researchers then met to negotiate their disagreements~\cite{forman2007qualitative}, which resulted in re-assigning categories to some functions and restructuring six categorizations (L3 $\longrightarrow$ PP5, S3 $\longrightarrow$ L3, S4 $\longrightarrow$ S3, S5 $\longrightarrow$ S4, S6 $\longrightarrow$ S5, S0 $\longrightarrow$ PP0) and removing the `Other' category as it wasn't indicating any meaningful grouping of functionality. The final categories of all observed functions are listed in Appendix~(\cref{appen_classfication}), and further definitions of each category and function are provided in the supplementary materials~\cite{supp2023semsemake}.

\textbf{Encoding notebook functionality.} 
We automated the transformation of each notebook into an encoded version using a mapping of function calls to categories.
For each line of code, we extracted the sequence of categories ($H_1$) based on their linear order.
Then, we looked at permutations and nested functions, for instance, $np.mean (accuracy (y_{test}, pred)$ performs a summary (ST1) on the output of a machine learning model (ML4). We defined the rule to infer processes from nested calls in innermost-first order ($H_2$). We also looked at cases where custom functions were defined in the notebook and set the rule to get the sequence of processes at the point where they are called ($H_3$). We can transform notebook code into a list of categories using these three heuristics ($H_1$ - $H_3$).
To uncover meaning patterns within the encoding, we must address questions like pattern initiation points and how to deal with overlapping lines of code.

To facilitate identifying patterns in encoded notebooks, we defined the following heuristics that provide structural definitions to patterns. Loading data (category L) in any form indicates the start of a part of code written for a specific purpose ($H_4$), and getting any form of output like visualizations (V) or of a model (ML4) indicates the end of a purpose ($H_5$). However, if a sub-sequence of processes repeats multiple times, they should be reviewed to see if the pattern is meaningful ($H_6$). In some cases, we found meaningful purpose in repeated processes. For example, notebooks with multiple `model and verify output' loops were trying to perform the model selection by hand picking. Finally, if adjacent lines of code performed the same function (e.g., six lines of code/cells all doing \texttt{sort()} on different data), they were all grouped into one process of formatting data (PP3), and the sequence of 6:PP3 was instead counted as 1:PP3 ($H_7$) as we were interested in the purpose a part of the code performs). Using these heuristics ($H_4$ - $H_7$), we process the notebooks into a cleaner encoded list of categories. (See \cref{fig:ds_functions} in the Appendix \cref{appen_classfication} for the functional categories.)

\textbf{Identifying code purpose from encoding patterns.} Three researchers qualitatively analyzed repeating sequences of categories to identify which ones suggest ``what the code is meant to do.''
These sequences capture a logical purpose behind why multiple lines of code were written. For instance, when adjacent cells performed multiple data transformations followed by checking the summary of data operations, these cells' purpose was marked as ``summary-based transformation''. These code purposes were part of a step in the notebooks (e.g., summary-based transformation is part of data wrangling).

Once the authors identified the patterns, we automated frequency counting of the patterns by identifying the location of each function in each notebook.
The script takes in the processed encoded notebooks and consults a dataset where different sequences of categories associated with code purpose are stored. For example, a sequence of \{ML1, ML2, ML3\} or \{ML1, ML3, ML4\} or \{ML1, ML2\} all indicate the associated lines of code were meant to build a generic ML model.
If the script detects any of these sequences in the encoded notebook, it shows the pattern's location within the notebook.

\outliner uses these detected patterns and the location of the patterns to organize the code cells into sections. It uses the purposes of the code associated with each pattern as a section title. For example, in \cref{fig:birdview}, The code in cells 50-51 are typical ML model defining and training activities, which relates to the section ``Generic Modelling'' as shown in \cref{table:patterns} in Appendix \cref{appen_code_purpose}, for each code purpose, we presented explanations of each code purpose (code sequences of each code purpose in supplementary~\cite{supp2023semsemake}). \outliner stops the clustering sections using header markdown cells as chapter titles.

The chapter descriptions (introductory paragraph) were constructed by combining any text in the notebook by the original data scientist. To create an introductory paragraph (like books) that contains a summary of the chapter, we combined text sentences within each chapter into a paragraph and started this with: ``In this chapter, the data scientist...''.

Note that the code purpose we identified is based on repeated patterns in code functionality found in the 44 notebooks. We don't claim that these patterns are exhaustive, and other patterns may arise if the analyzed notebook uses functions other than the 605 functions we categorized. However, users can continue to use our method to identify meaningful patterns in previously unseen cases through two straightforward modifications. First, they must categorize and add the new functions to the list mapping stored in the configuration file. After using the updated configuration file to map the notebooks into a list of codes, users can use the pattern recognizer script to locate where the patterns exist.

\subsection{Operationalize Design Features}

We used a design probe to operationalize design. Design probes are a valuable strategy used by the Human-Computer Interaction (HCI) and Design community to detect flaws and challenges early on, and learn about participants' behavior. These probes are a design strategy that involves asking open-ended questions, presenting challenges, and observing participants' responses to build systems that support human processes \cite{wallace2013making}. Although not standalone systems, probes require significant effort to build and put design at the heart of the process \cite{boehner2007hci}. They are well-suited to verify the usefulness of theory-driven efforts and detect flaws and challenges early on \cite{drosos2020wrex, kery2020mage}. Additionally, probes create an openness in the study to learn about participant behavior and provide an alternative approach to participatory design that starts with blank pages \cite{gaver1999design, gaver2001presence}.

We have named our design probe \Outliner, as it is specifically designed to parse the notebook's ``purpose'' and display it in interactive groups. At this stage, our methods are semi-automated, but we have the opportunity to improve automation and add intelligent parsing and rendering when the probe evolves into a full-scale system. There are two steps to transform a notebook into a \Outliner notebook---first, identify the patterns present in the notebook using the pattern mining script package (provided in supplementary~\cite{supp2023semsemake}). And then map those patterns as headers into the front end of the \Outliner overlay built using JavaScript. Once mapped, \Outliner allows users to interact with all the features, navigate across sections, annotate the notebook, and export the notebook or parts of it.

%% file: tables/design_mapping.tex
\begin{table}[thb]
\caption{Mapping Design Features and Cognitive Tasks in Sensemaking}
\label{table:mapping}
\resizebox{4.5 in}{!}{
\renewcommand{\arraystretch}{1.3}
\begin{tabular}{ll}
\rowcolor[HTML]{EFEFEF} 
\textbf{Cognitive tasks} & \textbf{Deisgn Features} \\ \hline
Understanding Behavior (C1) & \protect\overviewLabelBig{A} \protect\overviewLabelBig{B} \protect\overviewLabelBig{C} \protect\overviewLabelBig{D} \protect\overviewLabelBig{E} \\
\rowcolor[HTML]{EFEFEF} 
Comparing Information (C2) & \protect\overviewLabelBig{A} \protect\overviewLabelBig{D} \protect\overviewLabelBig{E} \\
Broadening Knowledge (C3) & \protect\overviewLabelBig{D} \protect\overviewLabelBig{E} \\
\rowcolor[HTML]{EFEFEF} 
Understanding Structure (C4) & \protect\overviewLabelBig{A} \protect\overviewLabelBig{B} \protect\overviewLabelBig{C} \protect\overviewLabelBig{D} \protect\overviewLabelBig{E} \\
Mapping/Relating Elements and Concepts (C5) & \protect\overviewLabelBig{A} \protect\overviewLabelBig{D} \protect\overviewLabelBig{E} \\
\rowcolor[HTML]{EFEFEF} 
Inferring hypothesis (C6) & \protect\overviewLabelBig{A} \protect\overviewLabelBig{D} \protect\overviewLabelBig{E} \\
Inferring elements (C7) & \protect\overviewLabelBig{A} \protect\overviewLabelBig{B} \protect\overviewLabelBig{D} \protect\overviewLabelBig{E} \\
\rowcolor[HTML]{EFEFEF} 
\multicolumn{2}{l}{\cellcolor[HTML]{EFEFEF}\protect\overviewLabelBig{A} \protect: Navigation panel; \protect\overviewLabelBig{B}\protect: Annotation; \protect\overviewLabelBig{C}\protect: Export; \protect\overviewLabelBig{D}\protect: Chapter title and icon; \protect\overviewLabelBig{E}\protect: Section title;}
\end{tabular}}
\Description{The Table 1 displays with different types of cognitive and design features. On the left side of the table, the cognitive tasks are listed, including Understanding Behavior (C1), Comparing Information (C2), Broadening Knowledge (C3), Understanding Structure  (C4), Mapping/Relating Elements and Concepts  (C5), Inferring hypothesis  (C6), and Inferring elements  (C7).

Each cognitive task is represented by a letter or letters. For example, Understanding Behavior is represented by the letter C1, Comparing Information is represented by C2, and so on. These letters are displayed in individual cells in the table.

On the right side of the table, the design features are listed. The letters A, B, C, D, and E represent the design features. These letters are also displayed in individual cells in the table.

At the bottom of the image, there is additional text describing the different elements represented by the letters A, B, C, D, and E. For example, A represents the Navigation panel, B represents Annotation, C represents Export, D represents the Chapter title and icon, and E means the Section title.}
\end{table}

%% file: design_probe.tex
\section{\Outliner Walk-through}
\label{sec:scenario}

\begin{figure}
\centering
\includegraphics[width=5in]{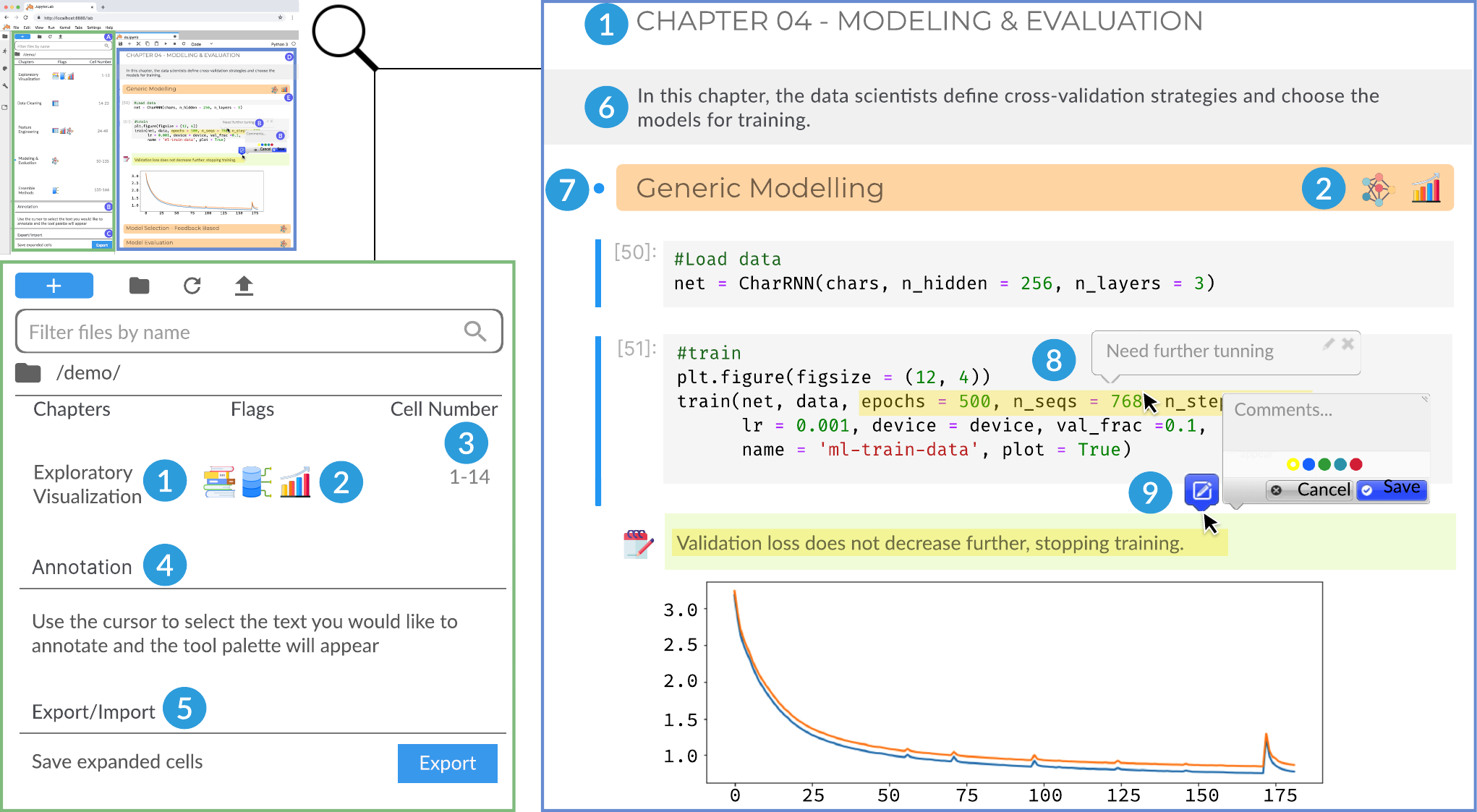}
\caption{Overview of Design Features}
\label{fig:panel}
\Description{The Figure 4 is a screenshot of the porpoise displaying in detail the feature description. 

In the top left of the image, there is a graph composed of smaller graphs, representing a small-scale version of the entire PORPOISE interface. A magnifier then zooms in to focus on both the left side panel and the main code sections on the right.

Below the small-scale plot, there is a zoomed-in view of the left side panel, which displays the panel in detail. This panel functions similarly to a table of contents. It provides a streamlined overview by showcasing chapter titles, allowing users to quickly identify major chapters such as "exploratory visualization," labeled as number 1. Enhancing its utility, the side panel indicates the number of cells associated with each chapter marked as number 3. As shown in the figure, the Ensemble method's code spans 1-14 cells. The side panel also features "flags" labeled as number 2 that highlight key operations within the notebook, such as the importing of data and libraries, the display of graphs and tables, and the building of models. Additionally, another feature displays annotations labeled as number 4. Below the annotation, a description acts as a description, stating, "Use the cursor to select the text you wish to annotate, and the tool palette will appear." Further below, there's a feature labeled as number 5, titled "export/import." The description under this title indicates "Save expanded cells," along with a blue export button.

On the right side of the small-scale plot, there is a zoomed-in view of the right expanding area. In the expanded Chapter 4 (Modeling & Evaluation), the content unfolds on the right-side canvas. This view displays the chapter title, labeled as number 1, followed by a brief introductory description of the chapter, labeled as number 6. The description reads, “In this chapter, data scientists define cross-validation strategies and select models for training.”

Below the chapter description, there's a section title labeled as number 7. Within each section headers area, "flags" are presented in the section headers, labeled as number 2. 

Below the section header, there are two code boxes. The first box is commented as 'load data', and the second box is labeled 'train'. Under these code boxes, there is a markdown comment. At the very bottom, there is a plot generated by the 'train' code. Within the 'train' code, there's a dialogue bubble labeled as number 8, indicating existing comments, which read: “needs further tuning.” Additionally, a cursor signifies that a user is currently commenting on the markdown sentence labeled as number 9, allowing the user to leave comments. This feature provides different color options for highlighting comments.
}
\vspace{-5mm}
\end{figure}

Charlie is a professional data scientist who uses computational notebooks in Python. She is exploring her team's past analyses to find an appropriate model for her current task. She uses \Outliner to explore the notebook (see \cref{fig:panel}).

\textbf{A bird's eye view:}
\Outliner provides a story-like overview about the notebook through the ``side panel'' (Green box in \cref{fig:panel}) that is similar to a table of contents. Charlie can see from the chapter titles (\cref{fig:panel}\overviewLabel{1}) that the notebook includes multiple chapters (e.g., Feature Engineering, modeling \& Evaluation), without having to read through the code. She also sees that the notebook uses ensemble methods, something she wants to use in her analyses. Therefore, she knows she is on the right path and decides to continue to explore the notebook. The side panel shows the amount of code per step, similar to chapter lengths, by displaying the associated cell numbers  (\cref{fig:panel}\overviewLabel{3}). Finally, ``flags'' in the side panel highlight the important operations like where data and libraries get imported, graphs and tables are displayed, and models are built (\cref{fig:panel}\overviewLabel{2}).

\textbf{Deep dive into areas of interest.} As Charlie wants to understand how the notebook implements the modeling, she ``opens'' Chapter 4 (Modelling \& Evaluation) by clicking the chapter name in the side panel to open it on the right side canvas (\cref{fig:panel}\overviewLabel{5}). The canvas shows the chapter title (\cref{fig:panel} \overviewLabel{1}) and a short introductory description of the chapter (\cref{fig:panel}\overviewLabel{6}). Charlie clicked the ``Generic Modelling'' section to see what is inside it (\cref{fig:panel}\overviewLabel{7}). Flags (\cref{fig:panel}\overviewLabel{2}) in the section headers provide cues about the modeling operations in each section, which helps her decide if the sections are of interest. So far, \Outliner has allowed Charlie to explore the notebook, its structure, and interesting operations without having to ``crawl'' through the code.

\textbf{Overlaying thoughts and comments.} As Charlie reviews the code by expanding each section, she formulates a few hypotheses she wants to test. \outliner allows data scientists to annotate parts of the notebook. Charlie jots down that the model  ``needs further tuning'' without having to actually modify the notebook (\cref{fig:panel}\overviewLabel{8}). Charlie selects a line of code, clicks on the annotate icon, which opens a comment (\cref{fig:panel}\overviewLabel{9}), where she writes her note and saves it. Notes are visible when a Charlie hovers over the annotated line of code (\cref{fig:panel}\overviewLabel{8}).
After exploring the relevant chapters and code cells, Charlie creates a snapshot of her exploration by exporting the notebook, which creates a document (PDF) (\cref{fig:panel}\overviewLabel{5}) with the expanded cells and annotations.

%% file: user_evaluation.tex
\section{User Evaluation of \Outliner}
\label{sec:sec5}

We conducted a within-subject counterbalanced observation study to learn how data scientists make sense of and explain notebooks using \Outliner's interaction experience against baseline Jupyter. To compare their experiences and understand how the adaptations of \Outliner help improve sensemaking, we structure our analysis to answer the research question:
\textbf{RQ2: How helpful are computational notebooks that support sensemaking?}

\subsection{User Study }

\input{tables/demo_user_study}

\emph{Recruitment.} We recruited 24 data science practitioners using convenience and snowball sampling from companies like Microsoft, RStudio, Amazon, and GitHub, and data science practitioners with advanced mathematics and computer science degrees. The participants were recruited through a sign-up survey. The Institutional Review Board (IRB) at our University approved the study protocol and materials and determined that the study had minimal risk. We compensated participants with \$50 gift cards for their time. Participants reported working in various industries and roles, including data analysts, data scientists, and machine learning developers (\cref{tab:userstudyparticipants}). They primarily reported using Python with Jupyter Notebooks, motivating them to participate in the study.

Once participants scheduled a time for a video call via Zoom, we forwarded the review board-approved consent form. The study was conducted entirely online through Zoom. All parts of the study were browser-based, and participants only needed access to a browser. The authors followed a script to introduce the study objectives and walked them through the parts of the task. Upon receiving verbal consent from the participants, the researcher recorded the voice and screen activity of the browser window.

The study was conducted during the COVID-19 pandemic when most people worked from home or remotely. We chose to make the study more accessible by selecting a browser-based approach. This lowered the barrier as participants did not need to install any apparatus on their devices.

\emph{Protocol.} We conducted a within-subject counterbalanced observation study, incorporating a think-aloud protocol to capture verbalized thoughts while performing a task \cite{fonteyn1993description}. We asked participants to go through two notebooks authored by two anonymous data scientists and ``explain the logic behind the analysis as [they] go through the code. Especially, focus on what [they] think the data scientist was trying to achieve or how they approached the problem'',  which we will refer to as the \textit{task} henceforth. We tentatively time-boxed each task to 25 minutes, and each participant performed the task twice, first explaining a notebook in Jupyter format followed by a notebook in \outliner format.

We selected Jupyter Notebooks since they are already ubiquitous in computational education and research among data scientists \cite{granger2021jupyter, perkel2018jupyter}. Notebook M is written to find the best model to predict movie ratings. Notebook H is an analysis to predict house pricing. The links to the source notebooks are provided in \cref{tab:userstudyparticipants}. The \outliner version of both notebooks is available here (redacted for anonymity and added to supplementary~\cite{supp2023semsemake}.)

For the first task, we provided Group 1 (P1-P12 in \cref{tab:userstudyparticipants}) with Jupyter Notebook \textit{M} hosted on GitHub and asked them to perform the task. At the end of this, we asked participants ``What are some of the pain points you faced when using this [Jupyter] format?''. Then we introduced the participants to \Outliner by demonstrating its features and allowing them to practice with a warm-up task. This involved expanding chapters and sections, highlighting code snippets, and leaving comments allowing them to familiarize with \Outliner. Once the warm-up was completed, the participants began the second task. The study protocol was the same for Group 2 (CP1-CP12 in \cref{tab:userstudyparticipants}). However, we switched the notebooks to counterbalance any patterns we observed from the notebooks themselves.

At the end of both tasks, we administered the participants with a post-study questionnaire~\cite{supp2023semsemake} asking them about the advantages and disadvantages of \outliner. Then, we asked six Likert scale questions where participants rated \outliner's helpfulness across different dimensions (See \cite{supp2023semsemake} for post-study questionnaire).

\emph{Analysis.} We first transcribed the video and audio recordings of the participants. From the first task, we gathered observations from the participants using Jupyter notebooks to comprehend the purpose of the code.  we analyzed the effect of using \Outliner. We performed a similar analysis using the verbalizations from participants and answers to the post-study questionnaires. The first three authors collaboratively mapped the participants' experiences to the broader themes of challenges discussed above, to present how these experiences changed when using \outliner. We also performed a descriptive analysis of the quantitative measures from the Likert scales to corroborate our findings.

\subsection{Effect of \outliner on sensemaking challenges}
\label{sec:porpoise_benefit}

\textbf{A. Helping users comprehend analysis structure and purpose.} \Outliner presents the notebook with chapters on the side panel, chapter and section titles, and chapter descriptions.

The expandable sections of code and output were labeled with the code's purpose to provide easier pathways to understand the components. Participants felt that chapter names helped make sense of code more efficiently by making \textsl{``each chapter's main purpose very clear"} [P7]. \textsl{``Within this chapter, it was really helpful in terms of figuring out what really was going on in [chapter name]"} [P1, CP4, CP6].

Section titles within the chapters reduced participants' efforts to understand the analysis components. Participants initially verified whether the section title is representative by reading the code. Once they felt confident, they would ``skip" sections that don't interest them [P3, P4, P11, CP6].\textsl{``I will see does it explain the title? If it coincides with my understanding then I will tend to continue and not go through the detail of the code"} [CP11].

The collapsible property makes the length of the script not a concern, as sections break long parts into headers that can be hidden [P3, P5, CP6]. Additionally, cell numbers in \Outliner's side panel also provide a sense of how big the sections are, which can \textsl{``give you an idea of how much time it may take you  to read the chapter"} [CP2].

We observed participants could more easily compare alternate implementation pathways, e.g., go to feature engineering to compare the different models included in the notebook [P1, P4, P5, P6, P7]. Chapter and section titles can help participants understand the functionality of specific libraries and functions, \textsl{``I have guidelines here, I know you are doing data cleaning, feature engineering, etc"} [P9].

Overall, participants reported that features \overviewLabelBig{A}, \overviewLabelBig{D}, \overviewLabelBig{E} helped them understand the analysis components and purpose of the code. \textsl{``The chapter, chapter titles, chapter summaries, and then the notes within each section really helped me navigate and understand what was in the notebook that I was looking at"} [P3].

\textbf{B. Helping notebook users find relevant information.} \Outliner's navigational affordances helped significantly reduce the need to scroll across long notebooks. Since the chapters and sections are collapsible, Participants could \textsl{``simply click the section \ldots I don't need to go up and go down"} [P9, P11, CP8]. It also helped participants select to \textsl{``look at what matters to [them] right now"} [P3, P5, P6, P7, CP3, CP5], and they could \textsl{``focus on the content that is interesting''} [P5, P9, P11, CP9] as the \Outliner affordances \textsl{``guided where to go"} [P8]. These features made it \textsl{``much easier to locate specific code"} [P11, CP10] which helped participants orient themselves in the notebook and avoid getting lost [P9].

Participants could look for a specific portion of data based on the section title \overviewLabelBig{E} [P4] or flags indicating what elements are included in the chapter and sections [P1, P4, P12]. An interesting strategy participants tool was to evaluate which sections are most valuable and expand those,\textsl{``I would go to a chapter and then go to the section that I think would provide the most information about what are we looking for, because like, I could skip like libraries \ldots this limited the amount of code that I had to search through to find the piece of information I was looking for"} [P3].

\Outliner provides several UI affordances to help data scientists skip irrelevant or uninteresting information. As they already knew \textsl{``what is included based on the title"} [P2, P3], the section headers describing the code purpose allowed participants to skip or \textsl{``gloss over"} [P11] some sections. The collapsible property of the section headers made it \textsl{``easier to navigate''} [P5] and allowed participants to \textsl{``easily jump around''} the different sections [P4, P11] [C5-I8]. 
\textsl{``it was nice to be able to, like, skip over some things, because I knew that they weren't necessary"} [P3].

We observed this in action as CP9 was thinking out loud while trying to understand the notebook and said, \textsl{``Let's go on to the data cleaning, but I don't want to see this anymore. I'll close that"} [CP9].

\Outliner's navigational panel \overviewLabelBig{A} and icons also helped to make the experience interactive chapters \overviewLabelBig{D} locate components (graphs, data, etc.). The overall clear demarcation in the analysis through chapters and sections made it \textsl{``clearly much more easier to navigate"} [CP6, CP7, CP8]. \textsl{``Since the menu is available to me on the left-hand side, I could easily fiddle around. So, the usability index piece of the interface was better"} [CP9].

Participants found that \textsl{``flags are quite useful"} [P4] as it points to where certain phases of the analysis happen. CP6 explained, \textsl{``it is helpful because you see, I can see the graph button, where the model has been written, and where there are tables, and where there are things related to data \ldots that would help significantly."}

As a consequence of these dynamic interactions and the enhanced ability to traverse through various sections of the notebook,
participants reported being able to make more accurate assumptions about the code's structure and purpose overall [P1, P2, P3, P4, P5, P6, P12, CP1, CP2, CP4, CP9, CP11].

\textbf{C. Helping users build mental models of the notebook's purpose.} 

\Outliner helped participants to build a map of the structure and gave the participants \textsl{``the big picture"} of what the data scientist was \overviewLabelBig{A} 
\overviewLabelBig{D} \overviewLabelBig{E} trying to do [P3, P9, CP1, CP2]. \textsl{``Using this notebook, for each expanded section, at least I know like, oh, this, this operation is within this scope"} [CP11].

Participants could build and maintain mental models of the notebook by mapping the structure to the flow; as CP2 described, his \outliner experience was \textsl{``really like reading a book"}. The chapters on the side panel helped to \textsl{``very clearly to show the flow of the analysis''} [P5, P10, P11] which in turn made it easy for the participants to understand how each component fits into the overall objective of the notebook [P1, CP2].

The ability to \textsl{``get a quick, top-down view''} and \textsl{``choose what you want to drill into"} [P12] helped connect the different components into chained events towards the larger objective [CP1]. Participants could also \textsl{``review back to make links between different sections to see the technical connection behind those sections"} [P9]. This top-down view also helped participants to not be distracted.
\textsl{``When it comes to some internal operation, for example, transformation verification, before this notebook, when I look at those portions of code, I will go to the wild, small details, but here I don't I lose track of the big, big picture"} [CP11].

\textbf{D. Allow notebook users to build and share explanations.} 
Participants could leave annotations to themselves about hypotheses or their inferences, marking where to come back later if they needed to jump to another section [P5] \overviewLabelBig{B}. \textsl{``I could track the code based on the sections, annotation is one of the best parts} [P9]. While making sense of the notebook, CP9 needed to review some part of the code from later in the notebook. Instead, she marked the place to revisit later and moved on; \textsl{``So maybe I'll put a comment here, see you later"} [CP9]. 

Participants enjoyed this feature as it was \textsl{``similar to what [one] would do in a while reading a PDF. [One] would like to annotate it's, it's the same for this. It is very helpful"} [CP6]. Participants wanted to use annotation for various purposes, from noting down small questions like \textsl{``is this a typo?"} [P6] to take detailed notes about steps in the pipeline to remind themselves [P5] to \textsl{``review back"} [P9]. Additionally, many participants agreed that commenting feature would be helpful in a collaborative setting [P3, CP10].\textsl{``I could also see the comment feature being really important if I was going to be handing my work products to another person or back to the data scientist. Then that would be helpful too."} [P2].

Participants envision they would \textsl{``leave comments for collaborators" } [P3, P6, P10], \textsl{``readers/managers could leave comments if they have any confusion" }[P6], and \textsl{``send this notebook to the senior, to check out the results, or to present the results"} [P4, CP1, CP2 ]. They also suggested annotations can get \textsl{``help from the person who wrote the script"} [P11].

Furthermore, \Outliner's annotation feature uses color highlights for annotated text/code, aiding participants in their reading and understanding.
Participants discussed various strategies to use these features, like using separate colors for separate purposes [P10] or using highlighting as markers to go find information [CP8].\textsl{``You can label it with different colors. So I can see different colors organized by the tag. Oh, yellow will be my understanding or like my personal notes. And blue could be your to-do list \ldots"} [P9].

Finally, participants also found \Outliner's features to export annotated and expanded notebooks useful \overviewLabelBig{B} \overviewLabelBig{C}. Especially in collaborative settings, \textsl{``it is really hard to share thoughts"} [P9, CP4]. Sharing the annotated notebook would help communicate thoughts while working together [P9].

In contrast to participants' wanting to organize and drop down their thoughts for later on paper or external notepads [P6, P9, P10, P11] when using Jupyter notebooks, participants found \Outliner's annotation feature \textsl{``very helpful to indicate something for [themselves]''} [P5, P3, P6, P9, P11, CP1, CP2, CP4, CP8, CP9, CP10].

\begin{figure}
\centering
\includegraphics[width=4.5 in]{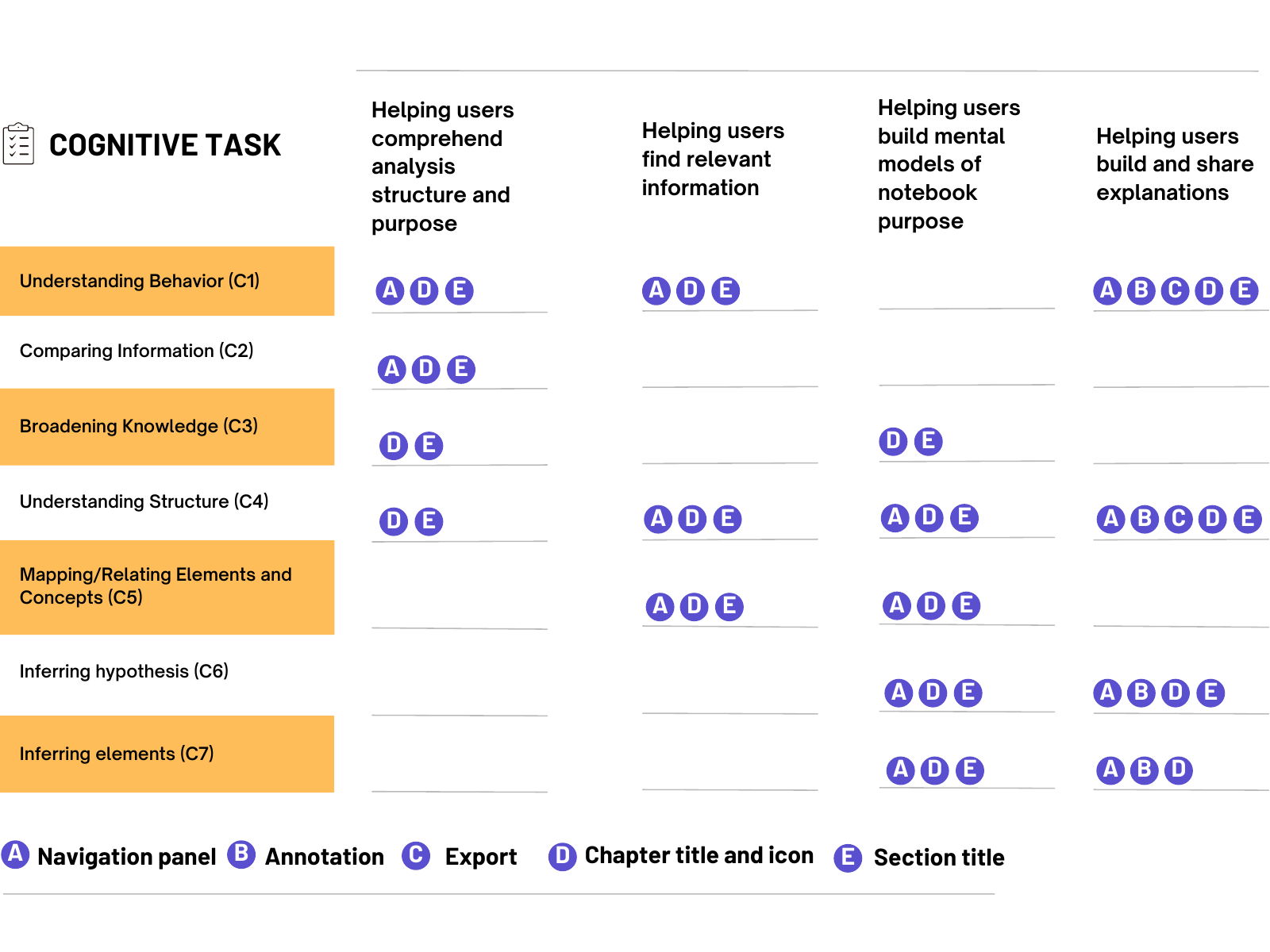}
\caption{Design feature mapping. Multiple features contribute to support each cognitive task in the sensemaking process.}
\vspace{-5mm}
\Description{The Figure 5 shows a table displaying multiple features that contribute to support each cognitive task in the sensemaking process. The table includes different types of tasks and challenges.

Above the challenges table, there is a section labeled "Cognitive Task”. Cognitive tasks are Understanding Behaviour (C1), Comparing information (C2), Broadening knowledge (C3), Understanding structure (C4), Mapping/Relating Elements and Concepts (C5), Interfering hypothesis (C6) and Interfering elements (C7). Challenges were “helping users analyze structure and purpose”, “helping users find relevant information”, “helping users build mental models of notebook purpose” and “helping users build and share explanations”

At the very bottom of the image, there is text for each of the letters in the former section. Letter A describes the navigation panel, B is the annotation, C is the export, D is the chapter title and icon, and finally, E is the section title.

It shows that each cognitive task has challenges in which part. For example, Understanding Behaviour (C1) has challenges in helping users analyze structure and purpose in A navigation panel, D Chapter title and icon, and E section title. That’s the same for Comparing information tasks but broadening knowledge and understanding structure only have this challenge in D and E. C1, C4 and C5 have challenges helping users find relevant information in A, D and E. }
\label{fig:mapping2}

\end{figure}

\textbf{An overview of mapping between design features and cognitive tasks.} To make it evident how the design feature assists the data scientist in better understanding unfamiliar notebooks. The mapping design features developed based on cognitive activities with sensemaking needs assessed in our user studies are shown in ~\cref{fig:mapping2}.
According to what we found, the cognitive tasks that are included in the comprehension element, the navigation panel, the chapter titles and icons, and the section titles are the three primary features that can assist data scientists in comprehending the objective, feature, and information of the analysis. As for structural-related cognitive tasks from the element of mental modeling, we found that in addition to the navigation panel, chapter/section titles, export, and annotations can provide additional help and support for data scientists who want to leave bookmarks for themselves and their collaborators. In addition, all of the features, with the exception of export, can assist data scientists in the process of inferring hypotheses and elements.

\begin{figure}
\centering
\includegraphics[width=4.5 in]{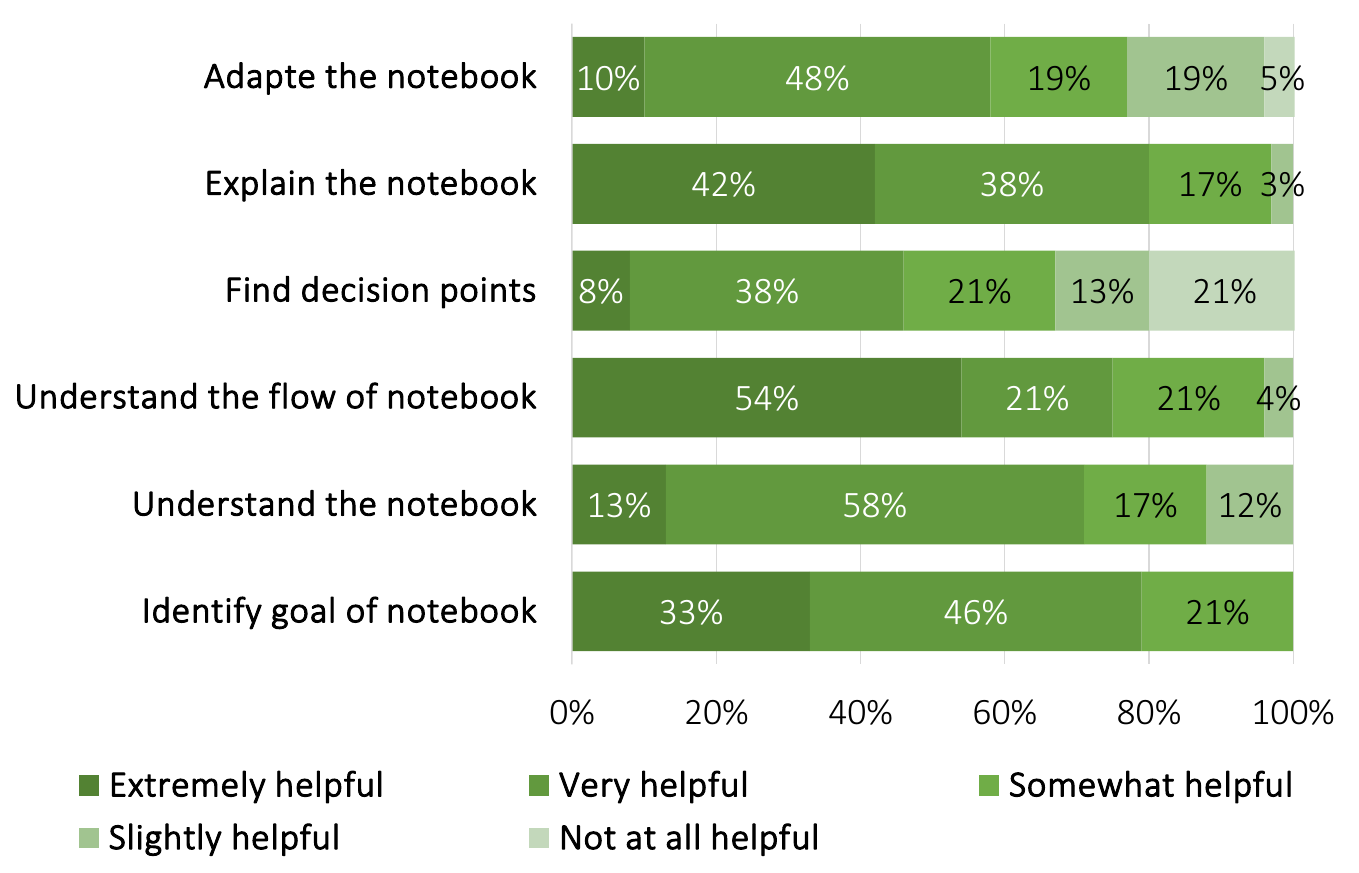}
\caption{The distribution of responses from six Likert-scale questions ranking the helpfulness of \outliner in making sense of notebooks, measured from 1 (Not helpful at all) to 5 (Extremely helpful). The y-axis presents the six dimensions along which \outliner's helpfulness is evaluated.}
\label{fig:likert}
\end{figure}

\textbf{Participants questionnaires.} 
\cref{fig:likert} shows the distribution of responses from participants about the helpfulness of \Outliner. We asked the participants to rate the helpfulness of \Outliner  by answering six questions. Four of these questions corresponded to cognitive tasks C1-C7, which are associated with three key elements in sensemaking: comprehension, mental modeling, and contextual inferences (see \cref{fig:sense_making}): ``Understanding the analysis'' [C1-C3] ``Identify the purpose of analysis'' [C1-C3], ``Understand the flow of the analysis" [C4, C5], and ``Find decision points'' [C6, C7]. The remaining two tasks, ``explain the notebook'' and ``adapt the notebook'', aren't the primary goals of our set of affordances; these are tasks that require sensemaking and are two of the possible activities that come after sensemaking. We used this set of questions to understand how \outliner's affordances help the sensemaking process and beyond and also to drive the discussions on how to improve their experience further. Participants ranked \outliner's ability to ``Understand the analysis'', ``Identify the purpose of analysis'', and ``Explain the analysis'' as the three most helpful aspects, with 71\%, 79\%, and 80\%,  of participants finding it very or extremely helpful.

Figure 6 displays a bar chart illustrating the distribution of responses from six Likert-scale questions. These questions rank the helpfulness of Porpoise in making sense of notebooks, with responses ranging from 1 (Not helpful at all) to 5 (Extremely helpful). The y-axis presents the six dimensions by which Porpoise's helpfulness is evaluated. The chart employs a 5-shade green color scheme. From left to right, the shades transition from darkest to lightest, representing the proportions of the 5 Likert scale responses received. At the chart's bottom, a legend specifies each option: "Extremely helpful", "Very helpful", "Somewhat helpful", "Slightly helpful", and "Not at all helpful", representing different levels of perceived helpfulness. The top section of the chart highlights the percentage of participants who agreed with the statement "Adapt the notebook." On the left, a displayed percentage reads "10\%." As one moves to the right, the percentages increase: "48\%", "19\%", and culminating in a combined "19\% and 5\%." The subsequent bar represents the percentage of participants who concurred with "Explain the notebook," accompanied by the respective percentages: "42\%", "38\%", and a combined "17\% and 3\%". These figures also seem to hint at the role of social media.
Next, another bar indicates the percentages of participants who agreed with the statement "Find decision points," with percentages displayed as 8\%, 38\%, 21\%, 13\%, and 21\%. Following that, another bar displays agreement on "Understand the flow of the notebook." The listed percentages from left to right are 54\%, 21\%, 21\%, and 4\%. Another bar then shows agreement levels on "Understand the notebook," with percentages of 13\%, 58\%, 17\%, and 12\%. The final bar showcases the percentage of participants who agreed with "Identify the goal of the notebook," with percentages displayed as 33\%, 46\%, and 21\%.

We discovered that the \Outliner is beneficial for data scientists to ``understand the flow of the analysis," with all participants finding \Outliner at least somewhat helpful. In comparison, almost a third of the participants did not find \Outliner helpful in finding decision points 34\%. Those who took part in the survey found \outliner ``very helpful to find the decision points'', as in the case of P2, who identified modeling and evaluations as the decision points. These opposing perceptions can be a result of how data scientists define decisions. While the purpose of code (shown in chapter and section titles on \outliner) based on \cref{table:patterns} is indicative of analysis decisions at a higher level of functionality, some data scientists might need more granular information to identify decision points (e.g., using $lineplot()$ instead of $boxplot()$). For example, CP3 asked \textsl{``what is the decision point''}.

24\% of the participants did not consider \outliner helpful for adapting the code (``slightly help \& Not at all helpful''). Several participants said they would need to \textsl{``develop [analysis] in this interface to answer this question''} [P2, P5, CP3]. They further elaborated that to evaluate adaptation, they would need to use \outliner on a notebook that specifically belonged to them [P5] or try adapting a notebook into their analysis: \textsl{``I am not sure, that was not the task, I did not do it''} [P2]. Whereas 58\% of participants believed that using an \Outliner will be at least somewhat beneficial in adapting and reusing notebook sections across different scales, for example, \textsl{``sections can serve as templates that be reused across the team''} [P3]. \textsl{``Because at least I don't have to go through the whole notebook and waste my time and then see how I can reuse or adapt''} [CP8].

\subsection{Design opportunities to further improve \Outliner}

At the end of the study, we asked participants to rate the helpfulness of the affordances for sensemaking. We asked six questions (See Figure~\ref{fig:likert}) to evaluate how well \outliner supports sensemaking and asked participants to elaborate on improving their experiences further. Following that, we engaged participants to discuss what support they ideally desire from their notebooks through the questions, ``What features of \Outliner do you find not helpful? What features would you like to have instead?".


\textbf{Adaptive definitions of Chapter and Sections.} Participants suggested adding the feature to interactively adjust the number of details in the chapter titles and section headers [P6, P9, P12]. Participants had different opinions about the level of detail to be captured in the chapter and section titles. For instance, P12 noted how it could sometimes be hard to ``tease out different results because they are very subtle'' and desired to be able to drill down into single code statement explanations. In other cases, P12 wanted coarser granularity in the section headers, for example, combining multiple sections into a single process-level section labeled ``Data Wrangling.'' Allowing users to choose the information captured in titles will provide personalized support for their individual sensemaking needs. Interfaces can allow users to add section or chapter titles and select which cells they want to include in each section.

\textbf{Making features interactive.}
For the side panel, we provided cell ranges to help data scientists understand the size of the chapters. However, none of the participants found the cell range useful and ignored it, with P3, P7, P11 calling it ``the least useful feature.'' In future implementation, presenting the number of cells within each chapter may be more useful. Participants generally found the flags for the structural components (data, library, graph, table, model) to be useful, and P5 and P12 suggested a capability that would allow them to add custom flags based on what information is relevant for specific tasks, for example, when a certain model is being used.

While highlighting the part of the code or text tied to the annotation is helpful to locate the comments, the \outliner provides ``just too many colors. And they that is a little bit distracting from the task at hand" [P2]. [P2, P5] further proposed adding search features to find and navigate between comments will improve usage; as P2 said, \textsl{``if there's a find, you can just go back and forth between the comments."}  Thus, future implementations can provide better support by highlighting the annotated parts with a single color, allowing users to navigate between comments, and allowing comments with different colors/icons to signify which user is commenting or replying to the comments.

%% file: tables/demo_user_study.tex
\begin{table}[thb!]
\caption{\label{tab:userstudyparticipants} Demographic Information of User Study Participants}
\resizebox{\textwidth}{!}{
\begin{tabular}{llllllll}
\rowcolor[HTML]{EFEFEF} 
\textbf{PID} & \textbf{Gender} & \textbf{Job Roles} & \multicolumn{1}{l}{\cellcolor[HTML]{EFEFEF}\textbf{DS Experiences}} & \textbf{PID} & \textbf{Gender} & \textbf{Job Roles} & \textbf{DS Experiences} \\ \hline
\multicolumn{4}{c}{\textbf{Group 1 (Jupyter notebook - M*; Porpoise - H*)}} & \multicolumn{4}{c}{\textbf{Group 2 (Jupyter notebook - H*; Porpoise - M*)}} \\ 
\rowcolor[HTML]{EFEFEF} 
P1 & Male & Computer science PhD student & \multicolumn{1}{l}{\cellcolor[HTML]{EFEFEF}3 - 5 years} & CP1 & Female & Computer science MS student & \textless{}1 year \\
P2 & Male & Data scientist/Staff researcher & \multicolumn{1}{l}{1 - 3 years} & CP2 & Female & AI/ML-engineer & \textless{}1 year \\
\rowcolor[HTML]{EFEFEF} 
P3 & Female & Data scientist & \multicolumn{1}{l}{\cellcolor[HTML]{EFEFEF}3 - 5 years} & CP3 & Male & AI/ML-engineer & \textless{}1 year \\
P4 & Male & Data scientist & \multicolumn{1}{l}{3 - 5 years} & CP4 & Female & Data scientist & 3 - 5 years \\
\rowcolor[HTML]{EFEFEF} 
P5 & Male & Computer science Post-doc & \multicolumn{1}{l}{\cellcolor[HTML]{EFEFEF}3 - 5 years} & CP5 & Male & Data scientist & 1 - 3 years \\
P6 & Male & Mathematics PhD student & \multicolumn{1}{l}{3 - 5 years} & CP6 & Male & Computer science PhD student & 3 - 5 years \\
\rowcolor[HTML]{EFEFEF} 
P7 & Male & Computer science PhD student & \multicolumn{1}{l}{\cellcolor[HTML]{EFEFEF}3 - 5 years} & CP7 & Male & AI-Ops/ML-Ops & 1 - 3 years \\
P8 & Male & AI/ML-engineer & \multicolumn{1}{l}{1 - 3 years} & CP8 & Female & Computer science PhD student & 1 - 3 years \\
\rowcolor[HTML]{EFEFEF} 
P9 & Male & Electrical engineering PhD student & \multicolumn{1}{l}{\cellcolor[HTML]{EFEFEF}3 - 5 years} & CP9 & Female & Computer science MS student & 1 - 3 years \\
P10 & Male & AI/ML-engineer & \multicolumn{1}{l}{1 - 3 years} & CP10 & Female & Data scientist/Staff researcher & 1 - 3 years \\
\rowcolor[HTML]{EFEFEF} 
P11 & Male & AI/ML-engineer & \multicolumn{1}{l}{\cellcolor[HTML]{EFEFEF}1 - 3 years} & CP11 & Female & Computer science PhD student & 3 - 5 years \\
P12 & Male & Data science researcher & \multicolumn{1}{l}{More than 5 years} & CP12 & Male & AI/ML-engineer & 1 - 3 years \\
\rowcolor[HTML]{EFEFEF} 
\multicolumn{8}{l}{\cellcolor[HTML]{EFEFEF}*M: Original source code: https://github.com/alexattia/Data-Science-Projects/blob/master/KaggleMovieRating/Exploration.ipynb} \\
\multicolumn{8}{l}{*H: Original source code: https://github.com/massquantity/Kaggle-HousePrices/blob/master/HousePrices\%20Kernel.ipynb}
\end{tabular}}
\Description{Table 2 shows the information related to the number of male and female participants in our counterbalance studies. The table is displayed on a white background.

The table consists of several columns, including "PID”,  “Gender”, “Job Roles" and "DS Experiences," which serve as headers for the different categories of data. The first column includes the identifiers "PID" and their respective values, while the second column contains the gender information of each individual, with entries for "Male" and "Female." The third column provides details about the job roles of the participant, such as "Computer science Ph.D. student" and "Data scientist/Staff researcher." The fourth column represents the "DS Experiences," indicating the number of years of experience in the field.

The table also includes two groups, "Group 1 (Jupyter notebook - M *; Porpoise - H*)" and "Group 2 (Jupyter notebook - H *; Porpoise - M*)." Each group is associated with specific assigned tasks.

Additionally, the image contains additional explanations below the table. These lines describe the sources of the original code used for reference in the projects mentioned in the table, with references to specific GitHub repositories.}
\end{table}

%% file: discussion.tex
\section{Limitations}
\label{sec6}
\reorg{\textbf{Automating \Outliner:}
By building \Outliner, we had the opportunity to evaluate the helpfulness of such a tool and provide researchers with promising directions for future research. \Outliner parses the notebook's purpose and displays it in interactive groups. While the parsing and the front-end display of \Outliner are implemented, \outliner can't automatically display parsed notebooks on the webpage, \cref{sec7} discusses avenues for automating \Outliner.}


\reorg{\textbf{Selecting notebooks:}
We made some assumptions about notebooks: we assumed that notebooks are linear and read top-down and that cells are executed in the order they are created. Our participants mostly use Jupyter notebooks, but environments substantially differ from notebook architectures---such as RStudio~\cite{rstudio} and Spyder~\cite{spyder}---may require different affordances than what \outliner provides.}

\changed{\textbf{Studying sensemaking individually:} Given the lack of understanding in how people make sense of data science artifacts, this paper focuses on how individuals make sense of notebooks authored by other data scientists. We discuss two features of Porpoise (annotation (B) and export (C)) in the context of collaborative data science.
In the future, we aim to conduct additional research to gain insights into data scientists' collaboration methods and to enhance \outliner for better support in social sensemaking.}

\changed{\textbf{User study:}
We conducted a counterbalance user study to investigate how data scientists use \outliner to understand and explain notebooks. When encountering unfamiliar interfaces, participants behave differently as they need time to situate themselves in the context. To reduce the unfamiliarity effect~\cite{russo1999unfamiliarity}, we evaluated \Outliner by first asking participants to use baseline Jupyter to situate them with their typical sensemaking strategies and behavior. Additionally, all user studies are limited by participants' response bias (as good-participant role~\cite{nichols2008good}).  To reduce this bias, we engaged participants to discuss negative experiences with \outliner, i.e., the least beneficial features for \outliner.}

\section{Discussion}
\label{sec7}



\changed{\textbf{Facilitating sensemaking in other domains.} Sensemaking is an essential part of doing tasks in any domain. Our findings about the benefits of \outliner and experiences with cases where \outliner lacked open exciting research opportunities to study how systems like this can support sensemaking in physical sciences and mathematics, literature, journalism, or even help make sense of other research papers. Since \outliner's features are based on fundamental psychological principles on how humans make sense of any knowledge artifact, researchers can study how similar support systems can be designed in their domains of interest.}

\textbf{Automatically elucidating code purpose to support code sensemaking and reuse.} 
The findings from our design probe are encouraging and suggest that the \Outliner interaction experience is useful to data scientists. Thus, we discuss approaches towards designing a robust mechanism for building a purpose catalog. There are multiple potential approaches to automatically generating this catalog.

One approach is to collect paraphrases of code comments and summaries describing code behavior from existing databases such as the MSR paraphrase corpus and phrase table~\cite{quirk2004monolingual}, ParaPhrase DataBase~\cite{ganitkevitch2013ppdb}, and DIRT~\cite{guichard2019assessing}. Additional code purposes can be added to the catalog using machine-translation-inspired generative approaches, or variational auto-encoders with sequence-to-sequence models~\cite{gupta2018deep} that can learn from existing sequences of code syntax using a feed forward Deep Neural Network.

Another approach is to use automated source code summarization~\cite{Haiduc2010automated} to describe the behavior of code cells using machine learning to discover the most important analysis steps~\cite{mcburney2015automatic}. Common methods to train machine learning models include collecting relevant keywords characterizing the program behavior using a Software Word Usage Model~\cite{hill2009automatically} and then using a Natural Language Generation system to generate natural language text describing the behavior~\cite{reiter1997building}. However, the same analysis step (and its code summarization) can mean different things based on the surrounding code. For instance, the same function when called in a data model selection step as compared to a model refinement step can have different purposes. Thus, automated code summaries must also take into account the context of a particular cell slice~\cite{Krinke06effectsof}.

\textbf{Using \Outliner to bootstrap literate programming.} While Jupyter notebooks have features like Markdown cells for data scientists to explain their notebooks, simply having the ability to do so doesn't mean that data scientists will actually provide meaningful descriptions~\cite{pimentel2019large}. While Knuth desired programs to be ``works of literature"~\cite{knuth1984literate}, even overcoming the initial inertia of having to document a notebook---especially an exploratory one---can be considerable.

\Outliner helps data scientists to overcome this inertia by automatically bootstrapping the notebook with descriptions. Just as revising paragraphs can be often easier than writing one from scratch, our participants remarked that being able to rename the automatic group descriptions---rather than having to come up with a description on on their own---made them more inclined to add descriptions (P4, P12).

For similar reasons, many participants reported the annotation capability as \textsl{``one of the best parts of the experience''} (P3, P5, P6, P8, P11). In contrast to explicit Markdown cells, \Outliner provided them a lightweight way to sprinkle their insights throughout the notebook. Consequently, the annotation experience was far less daunting when compared to more formal approaches to documenting the notebook. 

Finally, being able to reason about an unfamiliar notebook through the familiar experiences similar to physical books (P9) makes \textsl{``it easier to understand what to expect in each section through commonly understood terms like chapters and headings''} (P3). In addition to the self-explanation benefits of using \Outliner, we postulate that this interaction experience will also encourage data scientists to incrementally adopt literate programming practices in their own day-to-day notebook explorations.


%% file: conclusion.tex
\section{Conclusion}
\label{sec8}

\changed{From investigating sensemaking across psychology, computer science, and data science, we identified the components and cognitive tasks involved in sensemaking. We built a design probe of computational notebook overlay by blending five straightforward affordances that bring out the computational narratives by adding explanations about code structure and purpose. This design probe, \outliner, structures programs as explanations to humans rather than instructions to the computer. Through a  within-subject counterbalanced observation study with 24 data science practitioners, we found that even simple affordances to accentuate the computational narrative help cater to data scientists' challenges when sensemaking. Data scientists enjoyed \outliner as it helped them make sense of notebooks by getting a focused view and taking notes. P9 summarizes \outliner's benefits succinctly---\textsl{``It's just like reading a book!''}}

%% file: tables/Patterns_with_Icon.tex
\begin{table}[!thb]
\label{table:patterns}
\centering
\resizebox{5.5 in}{!}{
\renewcommand{\arraystretch}{1.3}
\begin{tabular}{lll}
\rowcolor[HTML]{EFEFEF} 
\multicolumn{2}{l}{\cellcolor[HTML]{EFEFEF}\textbf{Code Purpose}} & \textbf{Description} \\ \hline
\faArchive & Libraries & The process of configuring a programming system libraries. \\
\rowcolor[HTML]{EFEFEF} 
\faBuilding & system setup & Managing system resources such as files, directories, and environment variables \\
\faDatabase & Data loading & \begin{tabular}[c]{@{}l@{}}Loading multiple data sources or generating a population, and tuning the data\\ right away based on what they know about the data.\end{tabular} \\
\rowcolor[HTML]{EFEFEF} 
\faEject & Data generation & Making new data (quering database) \\
\faEject & Initial wrangling & Tuning the data based on internal knowledge/assumptions about the data. \\
\rowcolor[HTML]{EFEFEF} 
\faEject & Domain specific wrangling* & \begin{tabular}[c]{@{}l@{}}Loading and processing specific types of data like image, audio, geographical, \\ or text. Involves specific libraries and specific functional transformations.\end{tabular} \\
\faSave & Saving intermediate progress & Saving a model state with parameters, or saving transformed data for reuse. \\
\rowcolor[HTML]{EFEFEF} 
\faCamera & Visual Exploration of Data Space & Exploratory visual inspection of the data right after loading. \\
\faExchange & Data transform & \begin{tabular}[c]{@{}l@{}}Converting and modifying raw data into a structured format that is suitable for \\ analysis and modeling.\end{tabular} \\
\rowcolor[HTML]{EFEFEF} 
\faExchange & Data transform-Inspection based transformation & \begin{tabular}[c]{@{}l@{}}Data is processed/changed, inspected in textual or visual form, followed by \\ more change to the data.\end{tabular} \\
\faExchange & Data transform-Summary based transformation & Inspection of summary measures that leads to certain transformations. \\
\rowcolor[HTML]{EFEFEF} 
\faExchange & Data transform-Pre-model transformation & Doing data transformation to fit upcoming model \\
\faExchange & Data transformation verification & Building functions and transformations with interleaved data check along the way. \\
\rowcolor[HTML]{EFEFEF} 
\faExchange & Summary based transformation & Inspection of summary measures that leads to certain transformations. \\
\faEye & Pre-model inspection of data & \begin{tabular}[c]{@{}l@{}}Checking the shape and elements of data, in textual or graphical format, right \\ before feeding it into the model.\end{tabular} \\
\rowcolor[HTML]{EFEFEF} 
\faEye & Output verification & \begin{tabular}[c]{@{}l@{}}Multiple format inspection of a model's output or a transformation. Often also\\ involves checking the statistical summary.\end{tabular} \\
\faEye & Visual inspection & Looking at data/table/plot to inspect outpuit \\
\rowcolor[HTML]{EFEFEF} 
\faEye & Inspection based scientific coding & Building functions and transformations with interleaved data check along the way. \\
\faEye & Model output inspection & Using just a print of the metrics or the visualization of the model's output. \\
\rowcolor[HTML]{EFEFEF} 
\faCogs & Model selection-AutoML by hand & \begin{tabular}[c]{@{}l@{}}Selecting among multiple models. Typically either all models are initiated at the \\ same time, or they are run one after the other.\end{tabular} \\
\faCogs & Model selection-Feeback based & Select a different model based on the output metrics of the model. \\
\rowcolor[HTML]{EFEFEF} 
\faFlask & Model refinement & \begin{tabular}[c]{@{}l@{}}Improving the accuracy and performance of a machine learning model by adjusting \\ its parameters, optimizing hyperparameters, selecting better features, or trying out \\ different algorithms\end{tabular} \\
\faFlask & Model refinement-parameter turnning & \begin{tabular}[c]{@{}l@{}}When the same model is refined by changing the parameters, or model structure, \\ based on the result of the model's output.\end{tabular} \\
\rowcolor[HTML]{EFEFEF} 
\faFlask & Model refinement-Best practice & \begin{tabular}[c]{@{}l@{}}When standard processes are followed to tackle specific performance issues with \\ the model. These are handed down cookbook steps to take e.g. use ensemble to \\ reduce error in regression.\end{tabular} \\
\faFlask & Model refinement-Prior result based & Refining model based on previous results of the model \\
\rowcolor[HTML]{EFEFEF} 
\faFlask & Model refinement-Input tuning & \begin{tabular}[c]{@{}l@{}}Standard processing of data after finding data related issues causing model performance\\ to deteriorate\end{tabular} \\
\faMagic & Generic modeling & Typical cycle of define, train, test of an ML model. \\
\rowcolor[HTML]{EFEFEF} 
\faMagic & Modelling with defaults & \begin{tabular}[c]{@{}l@{}}Models based on the team/organization, or models that were pre-trained, or with default \\ parameters with no tuning/training\end{tabular} \\
\faPuzzlePiece & Statistical Modeling & Defining and executing a statistical model or performing statistical testing.
\end{tabular}}
\caption{Code Purpose from Functionality Patterns in Notebooks}
\Description{Table 3 is organized into two columns, the column headers include "Code Purpose", and "Description". The "Code Purpose" column provides a brief title of the purpose of the code. The "Description" column provides a more detailed explanation of the code's function.

Each row of the table corresponds to a specific code or process. The first row describes the process of configuring a programming system library. The second row pertains to managing system resources such as files, directories, and environment variables.

The third row involves data loading, including loading multiple data sources, generating a population, and tuning the data based on existing knowledge. The fourth row discusses data generation, specifically making new data or querying a database.

The fifth row focuses on initial data wrangling, which involves tuning the data based on internal knowledge or assumptions. The sixth row mentions domain-specific wrangling, which involves loading and processing specific types of data such as images, audio, geographical data, or text.

The seventh row mentions saving intermediate progress, which includes saving a model's state with parameters or saving transformed data for future use. The eighth row highlights visual exploration of the data space, which refers to exploratory visual inspection of the data immediately after loading.

The next rows discuss various data transformations and model-related processes. These include data transformation inspection, summary-based transformation, pre-model transformation, output verification, visual inspection, inspection-based scientific coding, model output inspection, model selection (both automated and feedback-based), model refinement, and standard model refinement processes.

The table also includes rows related to model refinement processes, such as parameter tuning, best practices, prior result-based refinement, input tuning, and generic modeling. Lastly, there is a row related to statistical modeling, which involves defining and executing statistical models or performing statistical testing.}
\end{table}